\documentclass[a4paper,11pt]{article}
\pdfoutput=1 % if your are submitting a pdflatex (i.e. if you have
\usepackage{jcappub} % for details on the use of the package, please
\usepackage{aas_macros}
\usepackage[T1]{fontenc} % if needed
\usepackage{mathtools}

\newcommand{\Quijote}{\textsc{Quijote} }
\newcommand{\Mpc}{\rm{Mpc}}
\newcommand{\hMpc}{h^{-1}\Mpc}

\newcommand{\de}{\text{d}}
\newcommand{\fnl}{f_{\rm NL}}
\usepackage{multirow}
\usepackage{caption}

\title{Probing primordial non-Gaussianity by reconstructing the initial conditions}

\author[a]{Xinyi Chen,}
\author[a,b]{Nikhil Padmanabhan,}
\author[c]{and Daniel J. Eisenstein}
\affiliation[a]{Department of Physics, Yale University, New Haven, CT 06511, USA}
\affiliation[b]{Department of Astronomy, Yale University, New Haven, CT 06511, USA}
\affiliation[c]{Harvard-Smithsonian Center for Astrophysics, 60 Garden St., Cambridge, MA 02138, USA}

\emailAdd{xinyi.chen@yale.edu}
\emailAdd{nikhil.padmanabhan@yale.edu}
\emailAdd{deisenstein@cfa.harvard.edu}

\abstract{
We propose to constrain the primordial (local-type) non-Gaussianity signal by
first reconstructing the initial density field to remove the late time
non-Gaussianities introduced by gravitational evolution. Our reconstruction
algorithm combines perturbation theory on large scales with a convolutional
neural network on small scales. We reconstruct the squared potential (that
sources the non-Gaussian signal) out to $k=0.2\ h$/Mpc to an accuracy of $99.8\%$. 
We cross-correlate this squared potential field with the
reconstructed density field and verify that this computationally inexpensive
estimator has the same information content as the full matter bispectrum. As a
proof of concept, our approach can yield up to a factor of three improvement in
the $f_{\rm NL}$ constraints over pre-reconstruction, although it does not yet include the complications of
galaxy bias or imperfections in the reconstruction. These potential improvements
make it a promising alternative to current approaches to constraining primordial
non-Gaussianity.

}

\begin{document}
\maketitle
\flushbottom

\section{Introduction}
The paradigm of inflation explains a few cosmological puzzles, but the underlying physics remains unknown. Over the years, there have been various inflation models proposed beyond the simplest ones, which predict Gaussian initial conditions. These more complicated models predict that the initial quantum fluctuations are non-Gaussian, at a level within reach by upcoming cosmological surveys \cite[e.g.][]{Pimentel22}. 
Information contained in primordial non-Gaussianity (PNG) can distinguish different inflation models. Current and next-generation large-scale structure (LSS) surveys and cosmic microwave background (CMB) experiments, such as the Dark Energy Spectroscopic Instrument \cite[DESI,][]{DESI16,DESI19}, {\it Euclid} \cite{Euclid13}, the Vera C. Rubin Observatory Legacy Survey of Space and Time \cite[LSST,][]{LSST12,LSST19}, the {\it Nancy Grace Roman Space Telescope} \cite{Roman15}, {\it SPHEREx} \cite{SPHEREx14,SPHEREx18}, MUltiplexed Survey Telescope \cite[MUST,][]{MUST}, Spec-S5 \cite{Spec-S5}, Simons Observatory \cite[SO,][]{SO19}, and CMB-S4 \cite{CMB-S4-science-book,CMB-S4-white-paper-21}, all have measuring PNG as part of their science goals to elucidate the nature of inflation and the primordial universe.

There are several broad classes of scale-invariant\footnote{as opposed to non-scale-invariant PNG, which are referred to as primordial features in the literature. These features are manifestations of dynamics in primordial universe (unsmoothed potential) and are typically oscillatory in power spectrum \cite{Pimentel22}.} PNG, resulting from different inflation mechanisms.
The local-type PNG is a sensitive probe of single-field and multi-field models. A value of $|f_{\rm NL}^{\rm loc}|\ll 1$ implies single-field inflation, whereas a large local PNG signal ($|f_{\rm NL}^{\rm loc}|\gtrsim1$) suggests the presence of additional fields beyond the inflaton (i.e. ``multi-field'').
Accordingly, the theoretically motivated sensitivity goal for the local type is $\sigma(f_{\rm NL}^{\rm loc})<1$. The local type is local in configuration space and peaks in the squeezed limit ($k_1\ll k_2\approx k_3$) triangle configuration. 
Non-local types of PNG are probed by equilateral ($k_1\approx k_2\approx k_3$) and
flattened ($k_1=k_2$, $k_3\approx 2k_1$) triangle configurations \cite{Senatore11, Cabass22b}.
A large signal of this type together with $|f_{\rm NL}^{\rm loc}|\ll 1$ indicates single-field non-slow-roll inflation, through self-interactions of the inflation field. Small signals in both local and non-local types of PNG indicate single-field slow-roll inflation \cite{Dore14,Pimentel22}. 
While these are the most commonly discussed parametrizations, other variants have been 
discussed in the literature as well \cite{Pimentel22}.

The current best constraint for the local $\fnl$ comes from the {\it Planck} CMB data and is  $f_{\rm NL}^{\rm loc}=-0.9\pm 5.1$ \cite{Planckfnl}. However, the primary CMB has limited modes to probe, due to its two-dimensional nature. Future experiments, including CMB-S4, will only achieve a factor of two improvement \cite{CMB-S4-science-book}, not enough to reach the aforementioned sensitivity goal. Including CMB secondary anisotropies, such as CMB lensing and CMB kinetic Sunyaev Zel'dovich (kSZ) effect, cross-correlated with LSS measurements, can potentially achieve $\sigma(\fnl^{\rm loc})=1$
\cite{SO19}. Our work focuses on using LSS measurements alone. The current best measurement using the LSS surveys and using power spectrum alone is 
$f_{\rm NL}^{\rm loc}=-3.6_{-9.1}^{+9.0}$ from the DESI DR1 quasar and luminous red galaxy samples \cite{Chaussidon24}.
This measurement will improve with more DESI data and other upcoming galaxy surveys. Unlike the CMB, LSS observations probe the three-dimensional density field and have access to more modes than the two-dimensional CMB. Upcoming LSS surveys are thus anticipated to set tighter constraints on 
PNG including achieving $\sigma(\fnl^{\rm loc})<1$ \cite{Dore14, Schlegel19,Sailer21}.

The state-of-the-art approach to measure $f_{\rm NL}^{\rm loc}$ in LSS is by constraining the scale-dependent halo bias in the power spectrum \cite[e.g.,][]{Chaussidon24,Mueller22,Castorina19,Ho15,Leistedt14}. 
Local PNG adds an additional large-scale linear bias in the form of $\Delta b=\fnl^{\rm loc}/(k^2T(k))$ at small $k$ \cite{Dalal08,Matarrese08,Slosar08}, where $T(k)$ is the matter transfer function. 
This approach only requires measuring the power spectrum, 
the simplicity of which makes it powerful. However, there are a number of systematic effects that impact such analyses. For example, the nonuniform density in the imaging data can result in spurious clustering on the large scales where $f_{\rm NL}^{\rm loc}$ dominates \cite[e.g.][]{Mueller22,Ho15}. 
Moreover, sample variance dominates at these scales, limiting the constraints on 
$f_{\rm NL}^{\rm loc}$, although this can be mitigated by multi-tracer analyses
\cite{Seljak09,Pimentel22,Yamauchi14,Ferramacho14,Alonso15,Fonseca15}.

To improve the constraints over power spectrum alone, it is useful to consider higher-order statistics \cite{Moradinezhad21}. A number of studies show that the bispectrum contains more PNG information and has significantly more constraining power for shot-noise limited samples \cite{Dore14,Ferraro19,Scoccimarro04,Sefusatti07,Baldauf11,Moradinezhad21, Barreira20,Heinrich23}. 
These improvements arise because the bispectrum of biased tracers probes both scale-dependent bias and the matter bispectrum, and the bispectrum can access more modes that contain PNG information compared to the power spectrum (by going to larger $k$). 
The bispectrum can also break parameter degeneracies in the power spectrum \cite[e.g.][]{Pimentel22,Coulton23}.

Progress has been made towards using the bispectrum to constrain $f_{\rm NL}^{\rm loc}$ \cite[e.g.][]{Cabass22,Damico22,Cabass22b}, but the level of improvement obtained over a power spectrum analysis varies across forecasts, simulations, and real survey measurements.
A joint analysis of the bispectrum and power spectrum was carried out by two studies on BOSS DR12 \cite{Cabass22,Damico22}, obtaining a 20\% improvement in the constraint on PNG.
Analyses of local PNG simulations with halo fields show between mild \cite{Coulton23b,Giri23,Jung24} 
to significant \cite{Moradinezhad21} improvement by adding the halo bispectrum. Several forecasts for upcoming surveys also show more significant improvements \cite{Ferraro19,Dore14,Heinrich23}. In summary, how much bispectrum improves $\fnl$ constraints seems to be dependent on multiple factors, such as scale cuts, sample number densities, volumes, biases, and priors.

There are challenges associated with using the bispectrum to constrain PNG. First, there are a large number of triangle configurations.
This large data vector together with its associated covariance matrix can be computationally expensive to measure. Further, gravitational nonlinearities and galaxy bias also induce a bispectrum, which dominates over any expected primordial signal and must be carefully modelled. 
In this paper, we tackle these two challenges with measuring the bispectrum.
We analyze a near-optimal compressed bispectrum estimator \cite{Schmittfull14} to mitigate the computational cost of calculating the full bispectrum. Our bispectrum estimator is optimized for local PNG, i.e. optimally extracting squeezed limit $\fnl$ information from the definition of the local type primordial potential. 
We also use density-field reconstruction techniques to estimate the primordial density and potential
fields and to significantly reduce the gravity-induced bispectrum. These techniques have primarily
been applied to baryon acoustic oscillation analyses, but as we demonstrate, can enhance the 
observed PNG signal. In this paper, we focus on the local-type PNG. So in what follows, we drop the ``loc'' superscript and use ``$\fnl$'' exclusively for local-type PNG.

This paper is structured as follows. Definitions and convention
are presented in Section~\ref{sec:theory}. We introduce the cross-power bispectrum estimator to constrain $\fnl$ and compare its constraining power to the bispectrum in Section~\ref{sec:near_optimal}.
We then focus on estimating the linear density and primordial potential fields required to measure the cross-power. Section~\ref{sec:sim} describes the simulations used in this study. Section~\ref{sec:phi2delta_recon} presents the reconstruction method and the performance, as well as the application of the cross-power to reconstructed fields. We also present forecasts for $\sigma(\fnl)$ in this section.
Section~\ref{sec:template} examines the discrepancy between reconstruction and the initial condition shown in Section~\ref{sec:phi2delta_recon} and the PNG information content in the field with template fits.
We discuss in Section~\ref{sec:discussion} and conclude in Section~\ref{sec:conclusion}.

\section{Theory}\label{sec:theory}

\subsection{Conventions}
We use the following Fourier transform convention:
\begin{equation}
\begin{split}
    \delta(\boldsymbol{k})&=\int \mathrm{d}\boldsymbol{x}
    e^{-i \boldsymbol{k}\cdot\boldsymbol{x}}\delta(\boldsymbol{x})\\
    \delta(\boldsymbol{x})&=\int \frac{\mathrm{d}\boldsymbol{k}}{(2\pi)^3} e^{i\boldsymbol{k}\cdot\boldsymbol{x}}\delta(\boldsymbol{k}).
    \end{split}
\end{equation}
The power spectrum is defined as 
\begin{equation}
\langle\delta(\boldsymbol{k})\delta(\boldsymbol{k'})\rangle=(2\pi)^3\delta_D(\boldsymbol{k}+\boldsymbol{k'})P(k),
\end{equation}
and the bispectrum of the density field as 
\begin{equation}\label{eq:density_bk}
\langle\delta(\boldsymbol{k}_1)\delta(\boldsymbol{k}_2)\delta(\boldsymbol{k}_3)\rangle=(2\pi)^3\delta_D(\boldsymbol{k}_1+\boldsymbol{k}_2+\boldsymbol{k}_3)B(k_1,k_2,k_3).
\end{equation}

\subsection{Local-type PNG}
The primordial potential with local-type primordial non-Gaussianity is conventionally characterized as follows (originally from \cite{Komatsu01}):
\begin{equation}\label{eq:local_potential}
    \Phi(\boldsymbol{x})=\phi_{\rm G}(\boldsymbol{x})+f_{\rm NL}\left(\phi_{\rm G}^2(\boldsymbol{x})-\langle\phi_{\rm G}^2\rangle\right)+...
\end{equation}
In this definition, $\phi_{\rm G}(\boldsymbol{x})$ is a Gaussian field. The nonlinearity parameter $f_{\rm NL}$ captures the deviations from Gaussian initial conditions.
Note that the field is squared in configuration space (a convolution in Fourier space).

In Fourier space, the primordial potential and linear density are related through
\begin{equation}\label{eq:delta_to_Phi}
    \Phi(\boldsymbol{k})=\frac{\delta_{\rm lin}(\boldsymbol{k})}{M_{\Phi}(k)},
\end{equation}
where 
\begin{equation}\label{eq:M}
    M_{\Phi}(k,z)=\frac{2}{3}\frac{k^2D(z)T(k)}{\Omega_{\rm m,0}H_{0}^2}.
\end{equation}
Here, $D(z)$ is the growth factor at redshift $z$, and $T(k)$ is the matter transfer function, $\Omega_{\rm m,0}$ is the matter density at the present day, and $H_{0}$ is the Hubble constant. The primordial bispectrum (of the potential $\Phi$) $B_{\Phi}$ and the density field bispectrum $B$ are related through (e.g. \cite{Taruya08})
\begin{equation}\label{eq:Phi_bk}
    B(k_1,k_2,k_3)=M_{\Phi}(k_1)M_{\Phi}(k_2)M_{\Phi}(k_3)B_{\Phi}(k_1,k_2,k_3).
\end{equation}
Note that we omit a subscript for statistics of the density field (here the bispectrum, and later in this paper also the power spectrum) for simplicity.
Using the above, the primordial bispectrum is given by
\begin{equation}\label{eq:primordial_bk_local}
    B_{\Phi}(k_1,k_2,k_3)=2\fnl \left[P_{\Phi}(k_1)P_{\Phi}(k_2)+2 {\rm\  perm.}\right],
\end{equation}
where $P_{\Phi}$ is the primordial power spectrum, most commonly parametrized by a power law, $P_{\Phi}(k)=A_s k^{n_s-4}$.

\section{Cross-power estimator for $\fnl$}\label{sec:near_optimal}

We propose to constrain $\fnl$ using the cross-power estimator, $\langle\Phi^2\delta\rangle$. Intuitively, 
this follows from the observation that $\fnl$ is the coefficient of $\phi_G^2 \approx \Phi^2$.
Indeed, \cite{Schmittfull14} showed that this estimator is
near-optimal\footnote{Here ``near-optimal'' refers to optimality in the limit of weak non-Gaussianity.}. 
Following \cite{Schmittfull14}, Sections~\ref{sec:multiple_parameters} and~\ref{sec:derive_cross_power}
reviews aspects of the derivation of
this general class of bispectrum estimators
and then focuses on the specific
case of $\fnl$. Section~\ref{sec:optimality} explicitly demonstrates that this $\fnl$ estimator contains the same 
information as the bispectrum.

\subsection{Bispectrum estimation for multiple non-Gaussian parameters}\label{sec:multiple_parameters}

We consider a bispectrum of the form
\begin{equation}
    B_{\rm th}=\sum_i f_i B_i,
\end{equation}
where the $B_i$ are fixed templates with unknown amplitudes $f_i$. One example would be Eq.~\ref{eq:primordial_bk_local}, where the corresponding $f_i$ is just $\fnl$; other examples 
are the growth, shift and tidal terms arising from gravitational evolution (see Section~\ref{sec:theory_PT} 
below), as well as the terms due to higher order galaxy bias and shot noise. To determine the $f_i$ from measurements
of $\delta(\boldsymbol{k}_1)\delta(\boldsymbol{k}_2)\delta(\boldsymbol{k}_3)$, we start from the log-likelihood
\begin{equation}\label{eq:bk_chi2}
    \chi^2= \int \frac{\de \boldsymbol{k}_1}{(2\pi)^3}\frac{\de \boldsymbol{k}_2}{(2\pi)^3}\frac{\de \boldsymbol{k}_3}{(2\pi)^3} (2\pi)^3\delta_{D}(\boldsymbol{k}_1+\boldsymbol{k}_2+\boldsymbol{k}_3)\frac{\left[\delta(\boldsymbol{k}_1)\delta(\boldsymbol{k}_2)\delta(\boldsymbol{k}_3)/V-\sum_if_iB_i(k_1,k_2,k_3)\right]^2}{\sigma_B(k_1,k_2,k_3)^2}\,,
\end{equation}
where $\sigma_B(k_1,k_2,k_3)^2$ is the variance of the bispectrum. Minimizing $\chi^2$ with respect to $f_j$ leads to solving 
a system of linear equations of the form $\sum_i\mathcal{M}_{ji}f_i=\theta_j$, where
\begin{equation}
    \mathcal{M}_{ji}=\int\frac{\de \boldsymbol{k}_1}{(2\pi)^3}\frac{\de \boldsymbol{k}_2}{(2\pi)^3}\frac{B_i(k_1,k_2,|\boldsymbol{k}_1-\boldsymbol{k}_2|)B_j(k_1,k_2,|\boldsymbol{k}_1-\boldsymbol{k}_2|)}{\sigma_B(k_1,k_2,|\boldsymbol{k}_1-\boldsymbol{k}_2|)^2}
\end{equation}
and
\begin{equation}\label{eq:theta_j}
    \theta_j=\int\frac{\de \boldsymbol{k}_1}{(2\pi)^3}\frac{\de \boldsymbol{k}_2}{(2\pi)^3}\frac{\delta(\boldsymbol{k}_1)\delta(\boldsymbol{k}_2)\delta(-\boldsymbol{k}_1-\boldsymbol{k}_2) B_j(k_1,k_2,|\boldsymbol{k}_1-\boldsymbol{k}_2|)}{V\sigma_B(k_1,k_2,|\boldsymbol{k}_1-\boldsymbol{k}_2|)^2}.
\end{equation}
For the rest of this paper, we will focus on the case of a single 
$f_i = \fnl = \theta_{\fnl}/\mathcal{M}_{\fnl, \fnl}$.
We discuss the impact of adding additional parameters in Appendix~\ref{appx:b2_fnl}.

\subsection{Cross-correlating $\Phi^2$ and $\delta$}\label{sec:derive_cross_power}

Following \cite{Schmittfull14}, we assume that the bispectrum is product-separable, i.e.
$B(k_1,k_2,k_3)=f(k_1)g(k_2)h(k_3)$, resulting in 
\begin{equation}
    \theta_j=\int\frac{\de \boldsymbol{k}_1}{(2\pi)^3V}\left[\int\frac{\de \boldsymbol{k}_2}{(2\pi)^3}\frac{f(\boldsymbol{k}_2)\delta(\boldsymbol{k}_2)}{P(k_2)}\frac{g(-\boldsymbol{k}_1-\boldsymbol{k}_2)\delta(-\boldsymbol{k}_1-\boldsymbol{k}_2)}{P(|-\boldsymbol{k}_1-\boldsymbol{k}_2|)}\right]\frac{h(\boldsymbol{k}_1)\delta(\boldsymbol{k}_1)}{P(k_1)}.
\end{equation}
A key insight in \cite{Schmittfull14} is that the term in the square bracket is a convolution 
in Fourier space and can therefore be rewritten as a product of two weighted density fields. Specializing 
to the case of $\fnl$, Eqn.~\ref{eq:Phi_bk} with $B_{\Phi^2}$ given in Eqn.~\ref{eq:primordial_bk_local} can be written as 
\begin{equation}\label{eq:density_bk_local}
    B_{{\rm th},\fnl}(k_1,k_2,k_3)=2\left(\frac{M_{\Phi}(k_3)P(k_1)P(k_2)}{M_{\Phi}(k_1)M_{\Phi}(k_2)}+2 \text{ perm.}\right) \,.
\end{equation}
This is clearly product-separable, and we obtain
\begin{equation}\label{eq:theta_fnl}
    \begin{aligned}
        \theta_{\fnl}&=6\int \frac{\de \boldsymbol{k}}{(2\pi)^3V}\frac{M_{\Phi}(k)}{P(k)}\left[\int \frac{\de \boldsymbol{k}_2}{(2\pi)^3}\Phi(\boldsymbol{k}_2)\Phi(-\boldsymbol{k}-\boldsymbol{k}_2)\right]\delta(-\boldsymbol{k})\\
        &=6 \int \frac{\de \boldsymbol{k}}{(2\pi)^3}\frac{M_{\Phi}(k)}{P(k)}{P}_{\Phi^2\delta}(k).
        \end{aligned}
\end{equation}
In the last equality, we introduce ${P}_{\Phi^2\delta}(k)$ defined by
\begin{equation}
(2\pi)^3\delta_D(\boldsymbol{k}+\boldsymbol{k'}){P}_{\Phi^2\delta}(k)= \langle\Phi^2(\boldsymbol{k})\delta(\boldsymbol{k'})\rangle.
\end{equation}
The term in the square bracket turns into $\Phi^2$ because convolution in Fourier space is a product 
in configuration space, i.e.
\begin{equation}\label{eq:convolution}
\begin{aligned}
\Phi^2(\boldsymbol{k})&=\int\mathrm{d}\boldsymbol{x}e^{-i\boldsymbol{k}\cdot\boldsymbol{x}}\Phi^2(\boldsymbol{x})\\
&=\int\mathrm{d}\boldsymbol{x}e^{-i\boldsymbol{k}\cdot\boldsymbol{x}}\frac{1}{(2\pi)^6}\int\mathrm{d}\boldsymbol{q} e^{i\boldsymbol{q}\cdot\boldsymbol{x}}\Phi(\boldsymbol{q})\int \mathrm{d}\boldsymbol{p} e^{i\boldsymbol{p}\cdot\boldsymbol{x}}\Phi(\boldsymbol{p})\\
&=\frac{1}{(2\pi)^3}\int \mathrm{d}\boldsymbol{q}\Phi(\boldsymbol{q})\Phi(\boldsymbol{k}-\boldsymbol{q}).\\
\end{aligned}
\end{equation}

This is our desired result, that $\fnl$ can be measured by cross-correlating the square of the 
primordial potential with the density field, $\langle \Phi^2 \delta\rangle$.
Note that \cite{Schmittfull14} propose $\langle\Phi^2\Phi\rangle$ as an estimator. These two forms are 
equivalent, since dividing the $\delta$ by $M_{\Phi}$ does not mix $k$ modes.
While we can further compress $P_{\Phi^2 \delta}$ into a single number $\theta_{\fnl}$, we 
choose to work with $P_{\Phi^2 \delta}$ as our fundamental observable.

Before proceeding, we note that the above expressions are modified in the 
presence of shot noise, although the overall structure is unchanged. We leave the 
study of the impact of shot noise for future work.

This estimator $\langle\Phi^2\delta\rangle$ can be expressed as an integral of the bispectrum,
\begin{equation}\label{eq:Phi2_delta_in_integral}
    \langle\Phi^2(\boldsymbol{k})\delta(\boldsymbol{k'})\rangle=\frac{1}{(2\pi)^3}\int\mathrm{d}\boldsymbol{q} M_{\Phi}(k')\langle\Phi(\boldsymbol{q})\Phi(\boldsymbol{k}-\boldsymbol{q})\Phi(\boldsymbol{k'})\rangle.
\end{equation}
The angle bracket term in the integrand is the primordial bispectrum. Importantly, we observe 
that the cross-power involves an integral over all $k$ modes, and is not naturally restricted to 
particular scales. This implies that a naive implementation of this estimator will require modeling 
the bispectrum out to high $k$. We return to this issue in the next section and later in the paper.

Finally, we note that this estimator is just one of a family (\cite{Schmittfull14}) of 
similar bispectrum estimators built from the 
cross-correlation of the product of two fields with a third field.

\subsection{Optimality of $\langle\Phi^2\delta\rangle$}\label{sec:optimality}

We now demonstrate that the cross-power spectrum has the same $\fnl$ information as the bispectrum. Ref.~\cite{Moradinezhad19} provides a forecast on $\fnl$ on an evolved galaxy field with a non-Gaussian initial condition comparing the cross-power and the bispectrum, finding the two statistics return the same information. Here, we focus on a linear density field, which is more relevant to our study.
Instead of using $\theta_{\fnl}$, we compare the cross-power spectrum directly to a model to determine $\fnl$.
The advantage of this approach is that comparing the shape of the cross-power spectrum to the 
theoretically expected shape could provide useful diagnostics for both systematic errors and 
unmodeled physical effects.
We use the 
Fisher formalism \cite{Fisher35,Tegmark97,Tegmark97b}, assuming data follow Gaussian distributions and the covariance is independent of the parameters of interest, and considering only one parameter, $\fnl$. We evaluate 
\begin{equation}\label{eq:Fisher}
    \sigma({\fnl})^{-2} = \boldsymbol{F}=(\partial\boldsymbol{y}/\partial \fnl)^{\intercal}\boldsymbol{C}^{-1}\left(\partial \boldsymbol{y}/\partial \fnl\right),
\end{equation}
where $\boldsymbol{y}$ is an array of our observable, either the bispectrum or the cross-power estimator $P_{\Phi^2\delta}$. 
We concentrate on the unmarginalized $\fnl$ error here, and defer a 
full treatment of degeneracies with other parameters to later work (but see also Appendix~\ref{appx:b2_fnl}).
We compute the Fisher errors about a fiducial 
Gaussian matter field with $\fnl=0$, using the same cosmology as the main analysis (see Section~\ref{sec:sim})
using the linear theory transfer functions and power spectra.

We start with the terms for the bispectrum. We assume the bispectrum takes the local shape, Eqn.~\ref{eq:density_bk_local}. The derivative is thus
\begin{equation}
    \frac{\partial B(k_1,k_2,k_3)}{\partial \fnl}=2\left(\frac{M_{\Phi}(k_3)P_{\rm lin}(k_1)P_{\rm lin}(k_2)}{M_{\Phi}(k_1)M_{\Phi}(k_2)}+2{\rm\ perm.}\right).
\end{equation}
Note that we use the density power spectra $P_{\rm lin}$ to emphasize that the observable is the matter 
density power spectrum.
The covariance matrix for the bispectrum is slightly involved, so we provide the details in Appendix~\ref{appx:bk_cov_short}.
The bispectrum Fisher matrix becomes
\begin{equation}
    F=\sum_{k_1=k_{\rm min}}^{k_{\rm max}}\sum_{k_2=k_{\rm min}}^{k_1}\sum_{k_3=k_{\rm min}}^{k_2}\frac{(\partial B/\partial \fnl)^2}{C^B}.
\end{equation}
As with calculating the bispectrum covariance matrix, we keep the triangle side length ordering as $k_1\geq k_2\geq k_3$.  
The summation must also satisfy the triangle rule that the third side is greater than or equal to 
(equal only for the folded triangle) the difference of the other two sides. 
Ref.~\cite{Scoccimarro04} has this expression in the more general form. 
We use $\Delta k=3\times k_f$ as our $k$-bin size.
where $k_f$ is the fundamental frequency. 

To calculate the derivative of $\langle\Phi^2\delta\rangle$, 
we use the fact that the cross-power estimator is an integral of the primordial bispectrum, Eqn.~\ref{eq:Phi2_delta_in_integral}. 
In terms of the primordial power spectrum, the derivative is 
\begin{equation}\label{eq:deriv_estimator}
    \begin{split}
\frac{\partial P_{\Phi^2\delta}(k)}{\partial\fnl}=2M_{\Phi}(k)\Bigg[ &\int \frac{\mathrm{d} \boldsymbol{k}_1}{(2\pi)^3} {P_{\Phi}(k_1)P_{\Phi}(|\boldsymbol{k}-\boldsymbol{k}_1|)}
+{2P_{\Phi}(k)}\int\frac{ \mathrm{d}\boldsymbol{k}_1}{(2\pi)^3} {P_{\Phi}(k_1)} \Bigg].\\
    \end{split}
\end{equation}
The first integral is just a convolution of the primordial power spectrum with itself
\footnote{The primordial power spectrum $P_{\Phi} \propto k^{n_s-4}$, where $n_s = 0.9624$ for 
our fiducial cosmology. For this particular choice of power law, both of the integrals above are divergent
at low $k$. We regulate this divergence by assuming 
$P_{\Phi}\propto (k+10^{-4})^{n_s-4}$ for the primordial power spectrum, and make a similar change 
to the bispectrum calculations. Note that, in both simulations and surveys, this divergence will be 
regulated by the finite volume which naturally introduces a low-$k$ cutoff.}.

We now turn to the covariance matrix for $P_{\Phi^2\delta}$, focusing on the disconnected terms
(Gaussian terms). We estimate $P_{\Phi^2\delta}$ as
\begin{equation}
    \hat{P}_{\Phi^2\delta}(k)=\frac{1}{N_{\rm modes}(k) VV_f}\int_{k} \mathrm{d}\boldsymbol{k}\Phi^2(\boldsymbol{k})\delta(-\boldsymbol{k}),
\end{equation}
where the integral is over all modes in a thin shell with radius between $[k-\Delta k/2,k+\Delta k/2]$. Here, $N_{\rm modes}$ is the number of modes in this thin shell: $N_{\rm modes}\approx4\pi k^2 \Delta k/V_f$ (up to order of $\Delta k$), where $V_f=k_f^3$ is the fundamental volumne and $\Delta k$ is the $k$-bin width. 
The covariance matrix for this estimator is 
\begin{equation}\label{eq:cov_phi2delta}
\begin{aligned}
    \boldsymbol{C}^{P_{\Phi^2\delta}}(k,k')&=\frac{2\delta_K(k,k')}{N_{\rm modes}(k)}P_{\Phi}(k)M_{\Phi}(k)^2\int\frac{\mathrm{d}\boldsymbol{k}_1}{(2\pi)^3}{P_{\Phi}(k_1)}{P_{\Phi}(|\boldsymbol{k}-\boldsymbol{k}_1|)}\\
    &+\frac{1}{\pi V}P_{\Phi}(k)P_{\Phi}(k')M_{\Phi}(k)M_{\Phi}(k')\int \de \Omega_{\hat{k'}} P_{\Phi}(|\boldsymbol{k'}-\boldsymbol{k}|).
    \end{aligned}
\end{equation}
The second integral is over the angle between $\boldsymbol{k}$ and $\boldsymbol{
k'}$\footnote{We assume that $\boldsymbol{k}$
points in the $\hat{\boldsymbol{z}}$ direction when doing these integrals.}. Note that the first term 
only contributes to the diagonal of the covariance matrix while the second term contributes to both the diagonal
and off-diagonal elements. However, the diagonal elements from the second term are subdominant so we 
drop these to avoid numerical complications.

In practice, we need to smooth the density/potential field before calculating the $\Phi^2$ term to avoid
coupling to smaller scale modes with lower fidelity. Including a smoothing function $W(k)$ modifies the above
expressions to
\begin{equation}\label{eq:deriv_estimator_smooth}
    \begin{split}
\frac{\partial P_{\Phi^2\delta}(k)}{\partial\fnl}=2M_{\Phi}(k)\Bigg[ &\int \frac{\mathrm{d} \boldsymbol{k}_1}{(2\pi)^3}W(k_1) {P_{\Phi}(k_1)W(|\boldsymbol{k}-\boldsymbol{k}_1|)P_{\Phi}(|\boldsymbol{k}-\boldsymbol{k}_1|)}\\
&+{2P_{\Phi}(k)}\int\frac{ \mathrm{d}\boldsymbol{k}_1}{(2\pi)^3} W(k_1)W(|\boldsymbol{k}-\boldsymbol{k}_1|){P_{\Phi}(k_1)} \Bigg]\\
    \end{split}
\end{equation}
and 
\begin{equation}
\begin{aligned}
    \boldsymbol{C}^{P_{\Phi^2\delta}}(k,k')&=\frac{2\delta_K(k,k')}{N_{\rm modes}(k)}P_{\Phi}(k)M_{\Phi}(k)^2\int\frac{\mathrm{d}\boldsymbol{k}_1}{(2\pi)^3}W(k_1)^2{P_{\Phi}(k_1)}W(|\boldsymbol{k}-\boldsymbol{k}_1|)^2{P_{\Phi}(|\boldsymbol{k}-\boldsymbol{k}_1|)}\\
    &+\frac{1}{\pi V}W(k)P_{\Phi}(k)W(k')P_{\Phi}(k')M_{\Phi}(k)M_{\Phi}(k')\\
    &\int \de \Omega_{\hat{k'}} W(|\boldsymbol{k'}-\boldsymbol{k}|)^2 P_{\Phi}(|\boldsymbol{k'}-\boldsymbol{k}|).
    \end{aligned}
\end{equation}

For the results below, we apply a cosine filter for $W(k)$, defined by
\begin{equation}\label{eq:cos}
  \Sigma_{\rm cos}(k) =
  \begin{cases}
    1 & \text{$k<k_{\rm min,cos}$} \\
   \cos\left[\frac{\pi(k-k_{\rm min,cos})}{2(k_{\rm max,cos}-k_{\rm min,cos})}\right]  & \text{$k_{\rm min,cos}\leq k\leq k_{\rm max,cos}$} \\
 0 & \text{$k>k_{\rm max,cos}$}
  \end{cases}
\end{equation}
This filter sharply suppresses the power in $\Phi^2$ for $k_{\rm min,cos} \le k \le2 k_{\rm max,cos}$ with no power at 
higher $k$.
In the following Fisher analysis, we apply cosine filters in three $k$ ranges: $[0.2, 0.25]\ h$/Mpc, $[0.4, 0.5]\ h$/Mpc, and $[0.8, 1]\ h$/Mpc. Given the above expressions, it is straightforward to calculate the 
Fisher matrix for the cross-power spectrum following Eqn.~\ref{eq:Fisher}.

\subsubsection{Results of the optimality test}

With the above calculations, we evaluate the Fisher error $\sigma(f_{\rm NL})$ as a function of $k_{\rm max}$ 
between $0.01-2\ \hMpc$. 
Note that the meaning of $k_{\rm max}$ is different in the two statistics. 
For the bispectrum, $k_{\rm max}$ represents the longest of the three sides, considering all triangle configurations up to this side length. In the case of cross-power, $k_{\rm max}$ is the scale at which 
the estimator is evaluated at. As Eqn.~\ref{eq:Phi2_delta_in_integral} indicates, the cross-power 
is sensitive to scales $k > k_{\rm max}$. Unfortunately, these are scales that are the most 
contaminated by nonlinear evolution and galaxy formation and are removed in any cosmological analysis. 
They are also where the reconstructed field deviates from the primordial field the most.
This is the reason why it is important to smooth the density field before computing $\Phi^2$, 
effectively band-limiting the density field at $k_{\rm cut}$.

\begin{figure}
    \centering
    \includegraphics[width=0.6\linewidth]{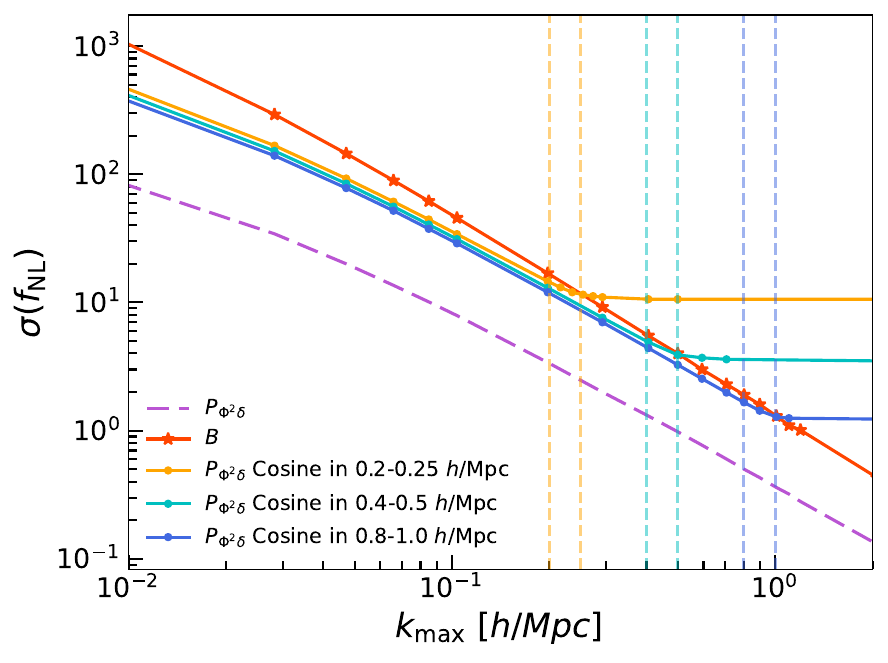}
    \caption{Fisher error $\sigma(f_{\rm NL})$ for cross-power estimator $P_{\Phi^2\delta}$ with a cosine filter in $k$ ranges of $0.2-0.25\ h/$Mpc (orange), $0.4-0.5\ h$/Mpc (cyan), and $0.8-1.0\ h$/Mpc (blue), and with no filter (purple dashed), compared to the full bispectrum $B$ (red), as a function of $k_{\rm max}$. 
    Note that $k_{\rm max}$ has different meanings for the two statistics (see text for more details). 
    We see that the errors in $\fnl$ from the cross-power estimator and the full bispectrum agree when all modes are 
    included (indicated by the second of each pair of vertical lines).
    }
    \label{fig:sigma_fnl_kmax}
\end{figure}

Figure~\ref{fig:sigma_fnl_kmax} compares the errors from the bispectrum and the cross-power estimator.
Without any smoothing, the errors from the cross-power spectrum are significantly smaller than the 
bispectrum, but this is an artifact of having $k_{\rm cut} \rightarrow \infty$. Introducing a band-limit 
using a cosine filter, we see that the errors are comparable. The dependence on $k_{\rm cut}$ is very weak, 
with the dominant contribution coming from the second term in Eqn.~\ref{eq:deriv_estimator_smooth} that 
grows approximately logarithmically with increasing $k_{\rm cut}$ for $n_s \sim 1$. We also observe that when 
we use the same $k$ modes, i.e. $k_{\rm cut}=k_{\rm max}$ 
(visually, at the second of the two vertical lines representing the 
cosine filter), both the bispectrum and cross-power spectrum yield the same errors.

\begin{figure}
    \centering
    \includegraphics[width=0.6\linewidth]{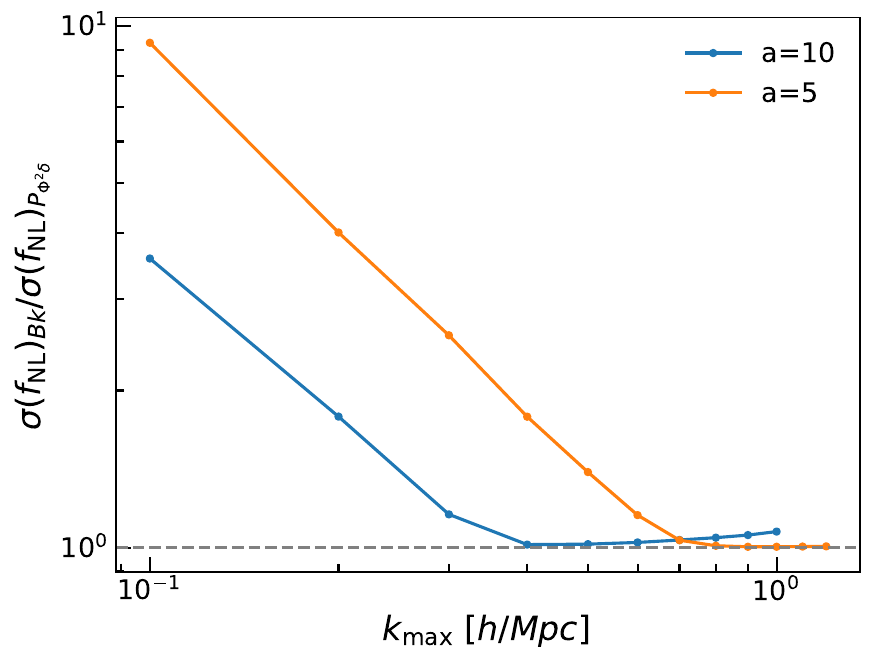}
    \caption{The ratio of $\sigma(\fnl)$ calculated from the bispectrum and from the cross-power for the 
    band-limited toy model $P_{\Phi}(k)=\exp(-a^2k^2)$ and $M_{\Phi}(k)=1$, for two choices 
    of $a$. As expected, when all the power has been integrated over, $\sigma(\fnl)$ from the cross-power
    equals that of the bispectrum. However, at lower $k$, the cross-power spectrum gives a lower error
    since it includes $k$ modes not explicitly included in the bispectrum (recall that $k_{\rm max}$ has a 
    different meaning for the bispectrum and the cross-power).
    The uptick in the blue line at high $k_{\rm max}$ is a numerical artifact.  }
    \label{fig:integrate_gaussian}
\end{figure}

To emphasize the fact that the cross-power spectrum is sensitive to all scales present in the data, 
and not just scales $< k_{\rm max}$, we consider a toy model where 
$P_{\Phi}(k)=\exp(-a^2k^2)$, and $M_{\Phi}=1$. Figure~\ref{fig:integrate_gaussian} demonstrates that the 
two errors in this case agree once we consider a large enough range of $k$ scales to account for all the power
in the model.

The advantage of analyzing the cross-power is in the computational cost. Assuming it takes $k_0$ number 
of bins to reach $k_{\rm max}$, the number of data points in the bispectrum scales as $k_0^3$, while a single cross-power spectrum data vector has dimension $k_0$. Even including multiple such vectors
to constrain other nuisance bispectra, the dimensionality would still only scale as $k_0$. This ensures
that these cross-power spectra will be more efficient to compute, while containing the same information 
as the full bispectrum.

\section{Simulations}\label{sec:sim}
We use the \textsc{Quijote} and $\Quijote$-PNG simulations \cite{Navarro20,Quijote-PNG,Quijote-PNG2} in this study. The \textsc{Quijote} simulations are a suite of $N$-body simulations with various cosmologies and resolutions. The $\Quijote$-PNG suite extends the $\Quijote$ suite to include PNG of different types in the initial conditions. Importantly, for the fiducial cosmology, the $\Quijote$ and $\Quijote$-PNG simulations have the same Gaussian initial conditions. For this work, we restrict ourselves to the local type PNG simulations. All of the simulations we consider are run in a 1 ($h^{-1}$Gpc)$^3$ box with 512$^3$ CDM particles and a background cosmology with $\Omega_{\rm m}=0.3175$, $\Omega_{\rm b}=0.0490$, $h=0.6711$, $n_s=0.9624$, $\sigma_8=0.8340$, $M_{\nu}=0.0$ eV, and $w=-1$. Additionally, we use a set of no-PNG simulations with the present day matter density $\Omega_{\rm m}$ changed by 3\% and all other cosmological parameters kept at the fiducial values. In addition to Gaussian initial conditions ($\fnl=0$), we use simulations with $\fnl=\pm100$. 
We use the initial conditions at $z=127$ and snapshots at $z=1$ and 0. We generate
density fields by distributing particles
on a $512^3$ grid using the triangular-shaped-cloud scheme (TSC, \cite{Hockney88}).
We ignore redshift space distortions throughout this study.

The initial conditions from the $\Quijote$ simulations are generated using a modified code of \cite{Scoccimarro12}. Essentially, the Gaussian primordial potential used to generate no-PNG simulations is modified to include $\fnl$ as described in Eqn~\ref{eq:local_potential}. 
The initial positions and velocities are generated with second-order 
Lagrangian perturbation theory (2LPT).
In our analysis, we use density fields derived from these positions to approximate 
the primordial density fields.
We will discuss the effect of this approximation in the later sections and
Appendix~\ref{appx:2lpt}.

\section{Application of $\langle\Phi^2\delta\rangle$ to reconstructed fields}\label{sec:phi2delta_recon}

To measure $\langle\Phi^2\delta\rangle$, we need to turn the observed late-time, low-redshift density field to a linear density field. We achieve this goal by applying a reconstruction technique to remove gravitational nonlinearity in the field.  The novelty of this current paper is in the application of this cross-power estimator to the reconstructed density and potential fields. 

In Section~\ref{sec:recon_method}, we describe our reconstruction method and its performance in terms of the direct output, the reconstructed density. We then consider the reconstructed squared-potential field in Section~\ref{sec:recon_phi2}. In Section~\ref{sec:phi2_delta_performance}, we 
present our cross-power statistic $\langle\Phi^2\delta\rangle$.
We forecast the expected $\fnl$ precision from this estimator 
in Section~\ref{sec:forecast}.

\subsection{Reconstruction method and performance}\label{sec:recon_method}
We use the hybrid reconstruction method developed in \cite{CNN} (a similar method can be found in \cite{Shallue23}) for this study. This method combines the strengths of perturbation theory on large scales and machine learning on small scales. The procedure to reconstruct the initial linear density field, $\delta_{\rm IC}$, from the nonlinear density field, $\delta_{\rm NL}$, involves two steps: 
\begin{enumerate}
\item We start with $\delta_{\rm NL}$ and first reconstruct with a perturbation-theory based traditional algorithm to obtain $\delta_{\rm rec, PT}$. 
\item We then train a neural network on $\delta_{\rm rec, PT}$ to produce $\delta_{\rm rec, CNN+PT}$, a density field closer to $\delta_{\rm IC}$. 
\end{enumerate}
This two-step process is motivated by the success of perturbation theory in describing matter distribution on large scales \cite[e.g.][]{Chen19,Desjacques18,Chudaykin20,dAmico20} and locality of convolutional neural networks (CNNs). 
In this study, we use the algorithm by Hada \& Eisenstein \cite{HE18} (hereafter HE18) for the pre-processing step. HE18 is an iterative reconstruction algorithm that aims to recover the linear density accurately in Lagrangian space. It features 2LPT and an annealing smoothing scheme, among other new ingredients compared to the standard reconstruction algorithm by \cite{Eisenstein07}. A detailed analysis of this method in comparison to the standard reconstruction algorithm can be found in \cite{recon}. 
We fix the effective smoothing scale to 10 $\hMpc$ and the iteration weight parameter 
to 0.5. The definitions of these parameters can be found in \cite{HE18,recon}.

For the CNN in the second stage, we use the same architecture presented in \cite{CNN}. The model has seven convolution layers and is designed to extract information in a 76 $(\hMpc)^3$ region 
to determine the reconstructed density at the center of that region. 
We train a model with eight simulations with $\fnl=0$ and test the model with two simulations also with $\fnl=0$. The remaining 90 simulations with $\fnl=0$ as well as the corresponding simulations with $\fnl=\pm100$ serve as the main samples for the analysis of this paper. The training is done by default on $\fnl=0$ simulations, even when the model is applied to $\fnl\neq 0$ simulations. We will discuss training variations later in the paper. We train two separate models for $z=1$ and 0. We focus on the $z=1$ case but comment on the $z=0$ case in some discussions. We denote the hybrid reconstruction as CNN+HE18.

We measure the performance of reconstruction by the residual power spectrum, defined as 
\begin{equation}
P_{\rm res}(k)=\left<|\delta_{\rm rec}(\boldsymbol{k})-\delta_{\rm IC}(\boldsymbol{k})|^2\right>/V,
\end{equation}
where $V$ is the volume and $\delta_{\rm rec}$ is the reconstructed density.
We also compute the propagator $G(k)$ and the cross-correlation coefficient $r(k)$ between the reconstructed field and the initial field, as these are commonly used in the literature.
The propagator is defined as
\begin{equation}
G(k)=\frac{\left<\delta_{\rm rec}^{*}(\boldsymbol{k})\delta_{\rm IC}(\boldsymbol{k})\right>}{\left<\delta^2_{\rm IC}(\boldsymbol{k})\right>},
\end{equation}
and the cross-correlation coefficient is defined as
\begin{equation}
r(k)=\frac{\left<\delta_{\rm rec}^{*}(\boldsymbol{k})\delta_{\rm IC}(\boldsymbol{k})\right>}{\sqrt{\left<\delta_{\rm rec}^2(\boldsymbol{k})\right>\left<\delta^2_{\rm IC}(\boldsymbol{k})\right>}}.
\end{equation}
Both these statistics measure the cross-correlation between the reconstructed density field and the linear density field; the difference is that the $G(k)$ also contains the amplitude information, whereas the $r(k)$ only has the phase information. If reconstruction were perfect, 
$G(k)$ and $r(k)$ both would be unity at all scales.

\begin{figure}
    \centering
    \includegraphics[width=0.6\columnwidth]{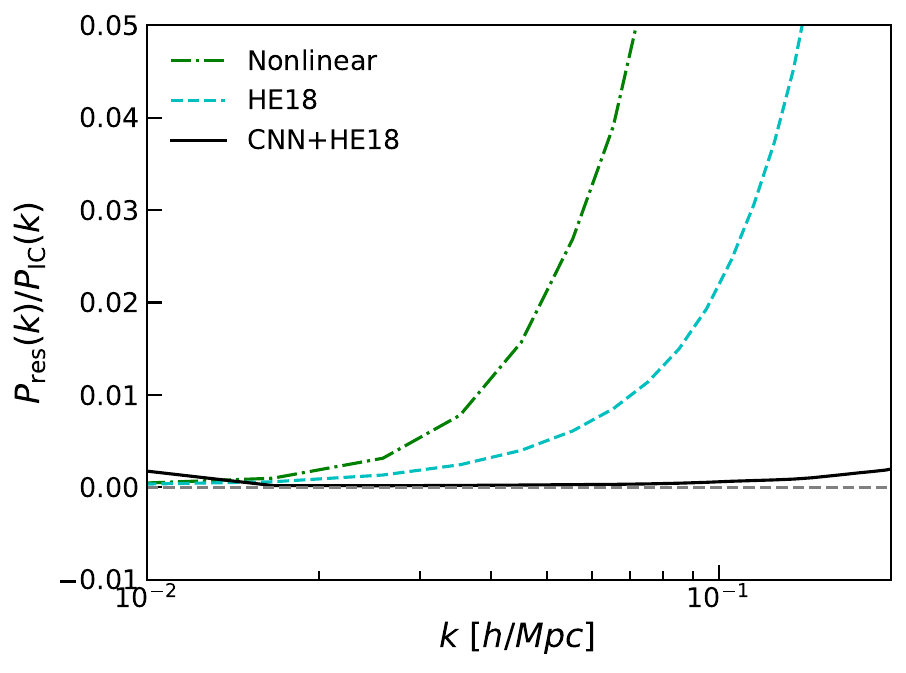}
    \caption{The performance of reconstruction of the density field measured by the residual power spectrum, comparing CNN+HE18 (black solid), HE18 (cyan dashed) and pre-reconstruction nonlinear (green dot-dashed) density fields at $z=1$ in real space. We show the performance up to $k=0.2\ h$/Mpc, which covers the $k$ scales used in the later analysis. We observe that CNN+HE18 matches the initial condition better than 0.2\% up to $k=0.2\ h$/Mpc.}
    \label{fig:recon_comp_CNN_HE18_NL}
\end{figure}

We now show the performance of this hybrid reconstruction in comparison with the HE18 reconstruction. This will be useful for the field-level analyses discussed later in the paper. Figure~\ref{fig:recon_comp_CNN_HE18_NL} shows the performance of CNN+HE18, HE18, and nonlinear density fields at $z=1$. The residual power spectrum of CNN+HE18 is below 0.2\% of the initial power spectrum up to $k=0.2\ h$/Mpc. The hybrid approach also achieves better than 0.4\% agreement in $G(k)$ and better than 0.1\% agreement in $r(k)$ out to $k=0.2\ h$/Mpc. A more detailed analysis of the performance can be found in \cite{CNN}.

\subsection{Reconstruction of $\Phi^2$}\label{sec:recon_phi2}

To explore why reconstruction can be beneficial, we analyze the contribution of  $\Phi(\boldsymbol{k})$ to the convolution in the $\Phi^2(\boldsymbol{k})$ field as a function of $k$-scale. 
We analyze the cross-correlation between $\Phi^2(\boldsymbol{k})$ with a $k$-cut and $\Phi^2(\boldsymbol{k})$ without a $k$-cut,
and we do this for a series of $k_{\rm max}$. In all cases, the cross-correlation coefficient drops to $\sim 0.75$ when $k=k_{\rm max}$, which suggests that the $k_{\rm max}$ controls where 75\% of cross-correlation coefficient can be achieved. We also find that roughly below $k_{\rm max}$/2, the cross-correlation is high and stays stable, but it drops quickly afterwards.
This suggests that the smaller scales one can do reconstruction to, the better $\Phi^2$ one can recover, improving the constraints of $\fnl$. This is different from the scale-dependent bias approach, where the signal is much stronger on large scales and diminishes as a function of $1/k^2$. In other words, smaller scales contain $\fnl$ information as well, and reconstruction helps to strengthen the signal from smaller scales.

Before discussing how well we can reconstruct the $\Phi_{\rm IC}^2$ field, we review how we estimate
it. We start by smoothing the reconstructed density field. This is necessary because 
squaring will 
couple the worse reconstructed short wavelength modes to the better reconstructed long wavelength modes.
We consider two smoothing schemes. 
The first is a Gaussian smoothing, with a smoothing kernel 
\begin{equation}\label{eq:Gaussian_smooth}
\Sigma_G(k)=\exp(-k^2R^2/2),
\end{equation}
where $R$ is the smoothing scale. We choose $R=5$ and 10 $h^{-1}$Mpc in this comparison. 
The second is a cosine filter introduced in Section~\ref{sec:optimality} in Eqn.~\ref{eq:cos}. With a hard $k$-cutoff, this scheme prevents mode coupling contribution from higher $k$-modes where reconstruction is less effective. In the following, we examine the case when $k_{\rm min,cos}=0.2\ h$/Mpc and $k_{\rm max,cos}=0.25\ h$/Mpc for the cosine filter.

Given the smoothed density field, we compute the potential using Eqn.~\ref{eq:delta_to_Phi}.
We then inverse Fourier transform back the potential to configuration space, square it and subtract the mean, 
and then Fourier transform the squared field to Fourier space.

We compare the $\Phi_{\rm rec}^2(\boldsymbol{k})$ field after reconstruction with the initial condition $\Phi_{\rm IC}^2(\boldsymbol{k})$ from the simulations. For the input density field, we consider the 
CNN+HE18 reconstruction along with the HE18 traditional algorithm as well as the nonlinear density field without any reconstruction. 
Similar to the density case, the residual power spectrum for the squared potential field is defined as
\begin{equation}
P_{\Phi^2,\rm res}(k)=\left<|\Phi^2_{\rm rec}(\boldsymbol{k})-\Phi^2_{\rm IC}(\boldsymbol{k})|^2\right>/V.
\end{equation}
In the above, both the reconstructed potential $\Phi_{\rm rec}$ and the initial potential $\Phi_{\rm IC}$ are smoothed with the same smoothing scheme and scale. We also look at both the corresponding propagator and cross-correlation coefficient.

\begin{figure}
    \centering
    \includegraphics[width=0.6\columnwidth]{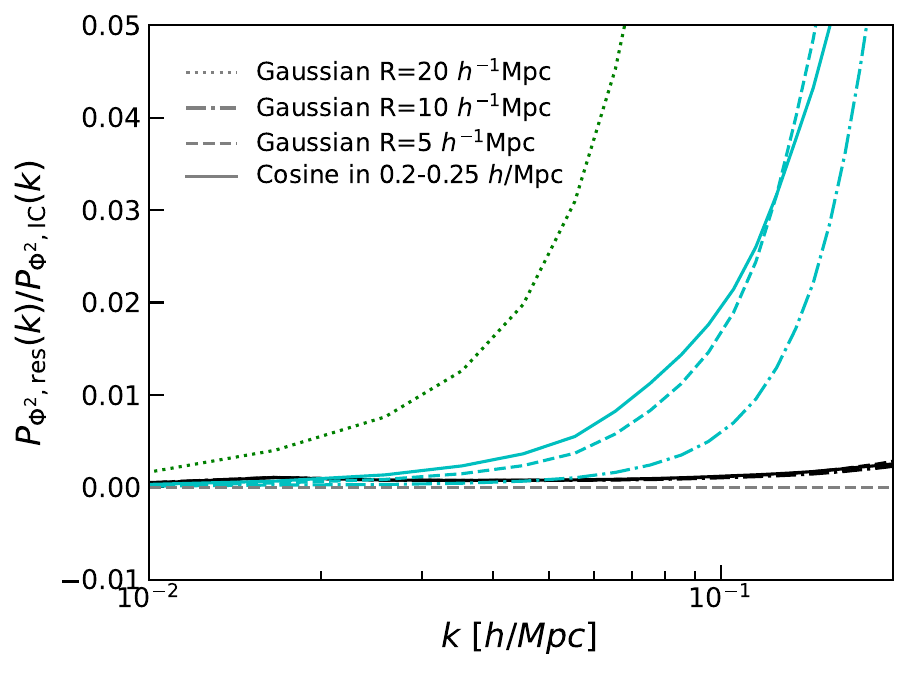}
    \caption{The performance of reconstructed field $\Phi_{\rm rec}^2$ measured by the residual power spectrum normalized by the initial power spectrum at $z=1$ in real space. We compare our CNN+HE18 model (black) with the traditional reconstruction algorithm HE18 (cyan) and with the nonlinear, pre-reconstruction field (green). We present several smoothing scales for Gaussian smoothing at 20 (only for pre-reconstruction, dotted), 10 (dash-dotted), and 5 (dashed) $h^{-1}$Mpc. We also apply a cosine filter (defined in Eqn~\ref{eq:cos}) in $k$ range $0.2-0.25\ h{\rm Mpc}^{-1}$ (solid). Both the nonlinear/reconstructed density and the initial density are smoothed. Each line in this figure is an average of 90 simulations that were not used in any stages of training. CNN+HE18 reconstructed field $\Phi_{\rm rec}^2$ closely follows the initial condition, with the power spectrum matching to the initial condition better than 0.3\% up to $k=0.2\ h{\rm Mpc}^{-1}$. In the hybrid reconstruction, all three lines are overlapping, suggesting that the
    cosine filter and Gaussian smoothing behave very similarly in this statistic at the scales shown here. }
    \label{fig:gk_rk_Phi2}
\end{figure}

Figure~\ref{fig:gk_rk_Phi2} shows the residual power spectrum of the reconstructed field, $\Phi_{\rm rec}^2(\boldsymbol{k})$, at $z=1$.
Each line is an average of 90 $\fnl=0$ simulations. The residual power spectrum $P_{\Phi,\rm res}^2(\boldsymbol{k})$ for the CNN+HE18 case is below 0.3\% of the initial power spectrum up to $k=0.2\ h$/Mpc. 
We find very little dependence on the smoothing scheme/scale in this two-point statistic.
The inferred $\Phi^2$ for the nonlinear field is very far from the initial, even when smoothed at 20 $\hMpc$, a relatively large smoothing scale. The $G_{\Phi^2}(k)$ and $r_{\Phi^2}(k)$ of the reconstructed field $\Phi_{\rm rec}^2(\boldsymbol{k})$ with the initial field $\Phi_{\rm IC}^2(\boldsymbol{k})$ are similar to the residual power spectrum.  The CNN+HE18 model closely follows the initial condition within 0.6\% for $G_{\Phi^2}(k)$ and 0.2\% for $r_{\Phi^2}(k)$ out to $k=0.2\ h{\rm Mpc}^{-1}$.

\subsection{Performance of $\langle\Phi^2\delta\rangle$ with reconstruction}\label{sec:phi2_delta_performance}
\begin{figure}
    \centering
    \includegraphics[width=\columnwidth]{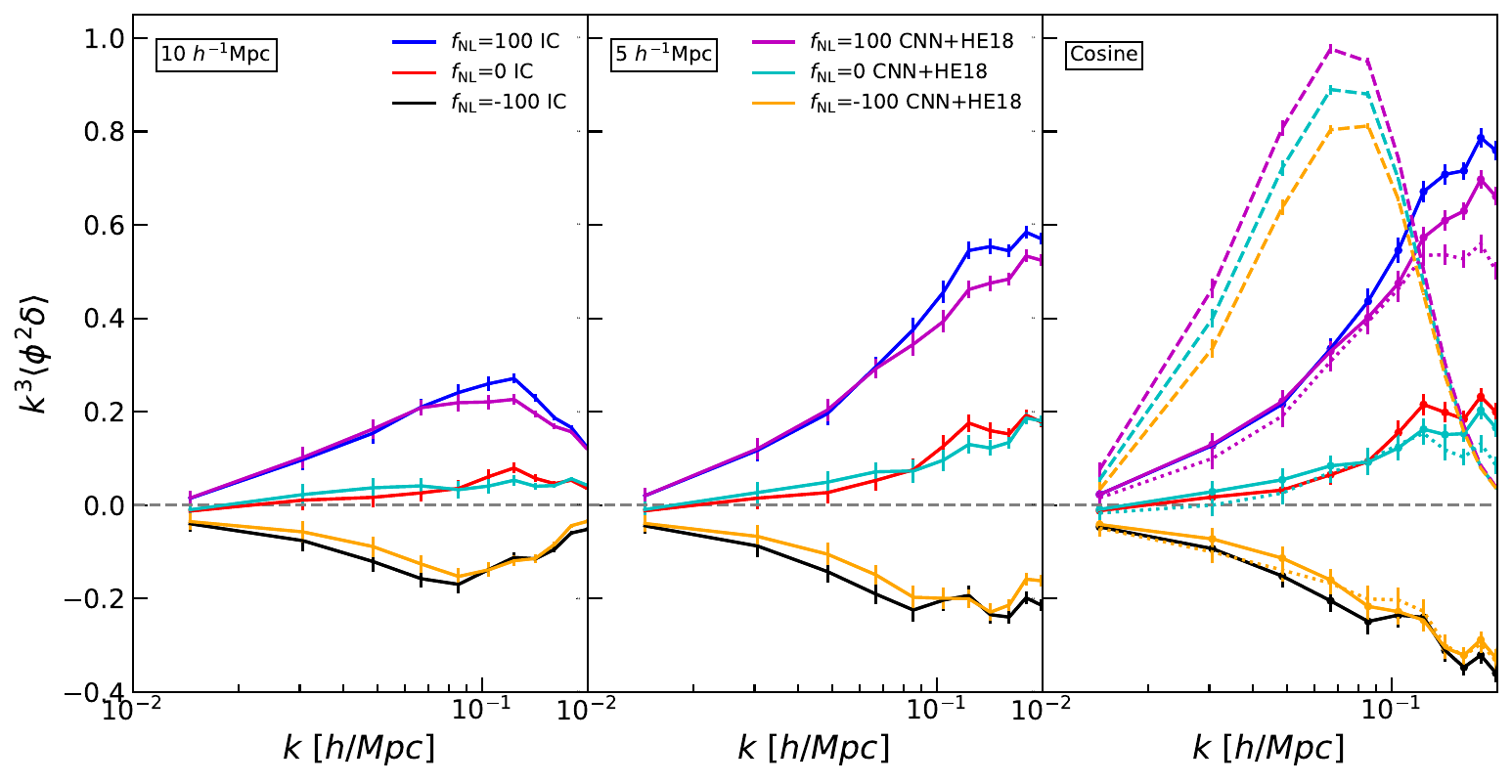}
       
    \caption{ The cross-power estimator $\langle\Phi^2\delta\rangle$ applied on the CNN+HE18 reconstructed fields and on the initial condition fields. The estimator is computed for $\fnl=100$ (magenta for reconstruction, blue for initial condition), $\fnl=0$ (cyan for reconstruction, red for initial condition), and $\fnl=-100$ (yellow for reconstruction, black for initial condition). We apply three smoothings for both reconstructed and IC density fields: Gaussian smoothing at $10\ h^{-1}$Mpc (left) and $5\ h^{-1}$Mpc (middle), and a cosine filter (defined in Eqn~\ref{eq:cos}) for $k$ in $0.2-0.25\ h{\rm Mpc}^{-1}$ (right). On the right panel, we also show the estimator applied on the corresponding pre-reconstruction fields at $z=1$ (smoothed at $R=20\ \hMpc$, dashed lines) and on reconstructed $z=0$ fields, which are rescaled to $z=1$ by the growth factor (dotted lines). The color scheme for these two follow that for the $z=1$ reconstruction cases. The error bars shown are errors on the mean.
    }
    \label{fig:statistic}
\end{figure}

We now show the estimator $\langle\Phi^2\delta\rangle$ applied to reconstructed $\Phi^2$ and $\delta$ fields and compare that to the estimator calculated with the initial condition fields. 
Figure~\ref{fig:statistic} shows the performance at $z=1$ with the aforementioned three smoothing schemes/scales: Gaussian smoothing with smoothing scale $R=5$ and 10 $\hMpc$, and cosine filter between $0.2-0.25\ \hMpc$. We plot $k^3\langle\Phi^2\delta\rangle$ such that the errors are about the same size on all scales. We also show one case of the cross-power calculated for the nonlinear field (i.e. pre-reconstruction) at $z=1$ as well as CNN+HE18 reconstruction at $z=0$ (scaled to $z=1$ by the growth factor) in the last panel for comparison. The amplitude difference in the three panels is due to different smoothings applied; 10 $\hMpc$ smoothing suppresses more power than 5 $\hMpc$ does and the cosine filter does not suppress power below $k=0.2\ h$/Mpc. The $\fnl=0$ statistic for the initial conditions is not at 0 due to 2LPT displacement in the initial condition (see Appendix~\ref{appx:2lpt} for details).

The constraining power of $\langle\Phi^2\delta\rangle$ is reflected as the signal-to-noise ratio, which qualitatively is the separation between the statistics at different $\fnl$ values over the error. 
The separation in the reconstruction versions is much larger compared to the nonlinear cases, although the errors are also larger. However, the increase in the signal is larger than the increase in the error,
thus reconstruction strengthens the PNG signal-to-noise in the field. The agreement with the initial condition in the estimator is better with $z=1$ than with $z=0$, while the errors are similar; therefore, reconstruction at $z=1$ has better constraining power. We also see that the differences among different $\fnl$ values in the nonlinear case are most prominent at low $k$, wearers these differences amplify out to higher $k$ after reconstruction. Thus, with reconstruction, more modes can be used for constraining $\fnl$.

We observe that there is discrepancy between the cross-power applied to the reconstructed fields and the cross-power applied to the initial condition fields. The discrepancy appears to be larger at higher $k$, due to reconstruction efficacy worsening at higher $k$.
The discrepancy is smallest with smoothing at 10 $\hMpc$.
Even though the power spectrum and cross-correlation between CNN+HE18 reconstruction and the initial condition for the $\Phi^2$ field as well as the $\delta$ field are better than 1\%, the discrepancy is still noticeable in our cross-power estimator. This is because the cross-power probes three-point statistic, which captures information not present in the power spectrum, and $G(k)$/$r(k)$ statistics, which are two-point statistics.

\subsection{Error estimate with Fisher forecast}\label{sec:forecast}
In this section, we perform a Fisher forecast for the $\fnl$ error from the $\langle\Phi^2\delta\rangle$ estimator. We calculate the Fisher matrix of the same form as shown in Eqn.~\ref{eq:Fisher}. However, instead of forecasting for the simplest setup of a Gaussian field with $\fnl=0$ as shown in Section~\ref{sec:optimality}, here we forecast for the actual $\Quijote$ simulations.
We measure $\langle\Phi^2\delta\rangle$ in 6 $k$-bins with $\Delta k=3\times
k_f$, where $k_f = 0.006\ h$/Mpc is the fundamental frequency. We use
the average $k$ in each $k$-bin, and we have $k_{\rm min}=0.015\ h$/Mpc and
$k_{\rm max}=0.1\ h$/Mpc. We show a few cases with $k_{\rm max}=0.2\ h$/Mpc as
well for comparison. To avoid the mismatches with a theoretical
model here (we revisit this in the next section), we use
the mean of the 90 simulations as the model.
The derivative is calculated as the difference in the cross-power between
$\fnl=\pm100$ simulations divided by 200.
We calculate the covariance matrix using the 90 $\fnl=0$ simulations, which is sufficient for 6 $k$-bins. We compare the diagonal elements
with the analytic estimate from Section~\ref{sec:optimality} and 
find reasonable agreement.

We perform a single parameter forecast for $\fnl$. As will be discussed in Section~\ref{sec:fitting_results} and shown in Figure~\ref{fig:degeneracy}, $\fnl$ is degenerate with the quadratic bias parameters in the model.
A single parameter forecast will thus underestimate the error in $\fnl$. 
However, our goal here is not to obtain the most accurate forecast, but to show that reconstruction can lead to a significant improvement in $\fnl$ constraints. We defer a multi-parameter forecast to our subsequent work, together with the application to galaxy fields, which requires more complicated modeling.

In Table~\ref{tab:forecast} we summarize the forecast for $\fnl$ for 
several cases.
As a limiting case, we also consider the Fisher error estimated assuming a perfect reconstruction, represented by the initial condition. We compare the reconstruction cases with the corresponding initial conditions smoothed at the same smoothing scales. The nonlinear case is smoothed at 20 $\hMpc$ only. 
We observe that the cosine filter results in tighter constraints than the Gaussian smoothing cases, which suggests that the Gaussian smoothing removes some primordial information on large scales. 
Using cosine filter comes closest to agreeing with the initial condition. 
We find that errors with higher redshift and larger $k_{\rm max}$ are lower. With higher redshift, reconstruction performs better, which confirms that removing gravity-induced nonlinearity strengthens the PNG signal and signal-to-noise. 

For the nonlinear field, the error at $k_{\rm max}=0.1\ h$/Mpc is about 100. Forecasts for the bispectrum using the same $\Quijote$-PNG simulations report an unmarginalized error in $\fnl$ of about 130 for $k_{\rm max}=0.1\ h$/Mpc (Figure 6 of \cite{Coulton23}), so our result is comparable but slightly better. Crucially, the errors on $\fnl$ are 
significantly larger before reconstruction than after reconstruction.
With the cosine filter and at $z=1$, we achieve a factor of 1.5 improvement in the $\fnl$ error estimate for $k_{\rm max}=0.1\ h$/Mpc and a factor of 3 for $k_{\rm max}=0.2\ h$/Mpc. However, we stress that this forecast is unmarginalized and the degeneracy between $\fnl$ and other parameters (discussed in Section~\ref{sec:template}) will increase
these error estimates.

\begin{figure}
    \centering
    \includegraphics[width=0.6\columnwidth]{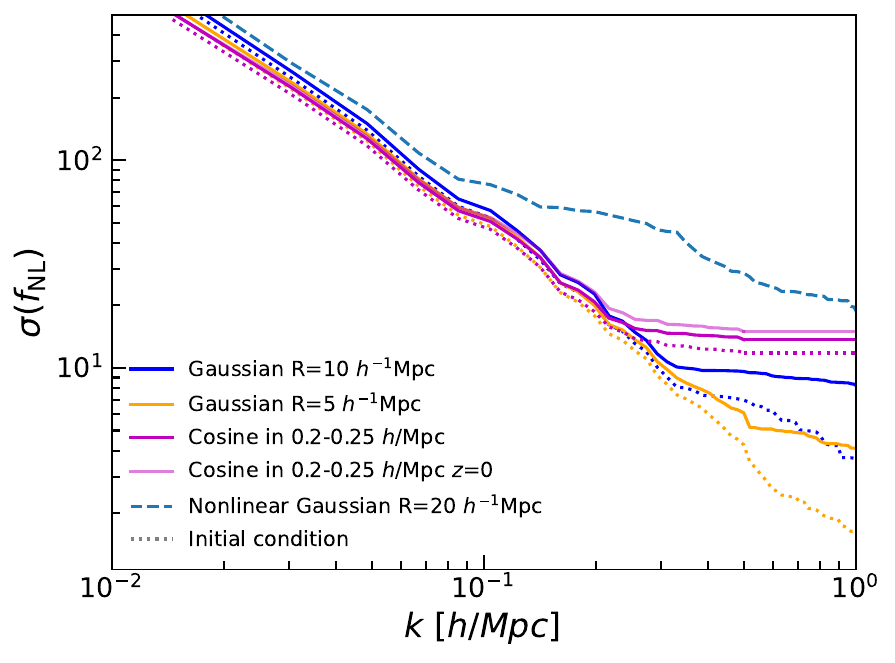}
    \caption{Forecast of $\fnl$ from the cross-power $\langle\Phi^2\delta\rangle$ computed on reconstructed fields with three smoothing schemes and scales at $z=1$. The three smoothings are Gaussian with scale $R=10\ \hMpc$ (light blue) and $R=5\ \hMpc$ (yellow), and cosine filter in $0.2-0.25\ h/$Mpc (dark magenta for $z=1$, light magenta for $z=0$). We also show the estimated error of the estimator computed on the unreconstructed field smoothed at $R=20\ \hMpc$ (dashed dark blue) and initial condition (dotted with colors as above) for all three smoothings for comparison. 
    The error from the reconstructed field agrees well with that from the initial conditions out to 
    $k \sim 0.3\ h$/Mpc, and is significantly smaller than the error from the nonlinear field.
    }
    \label{fig:sigma_fnl}
\end{figure}

Figure~\ref{fig:sigma_fnl} shows the cumulative effect of $\sigma(\fnl)$ as a function of $k$ for the same smoothing cases. 
The cosine cases slow down in decrease of the error at $k\sim 0.25\ h$/Mpc and flatten out at $k\sim 0.5\ h$/Mpc by construction. 
We could achieve smaller errors by including higher-$k$ modes (as 
is the case for the Gaussian filter). However, at these scales,
the disagreement between the estimator calculated with the reconstructed
and initial condition fields is also larger. We therefore 
choose to be conservative here 
and do not examine scales beyond $k_{\rm max}=0.2\ h$/Mpc.

\begin{table}[]
\centering
\caption*{\textbf{Fisher forecast for $\sigma(\fnl)$}}
\begin{tabular}{ccccccc}
\hline
                 $k_{\rm max}$& Smoothing & IC   & \begin{tabular}[c]{@{}c@{}}CNN+HE18  $z=0$\end{tabular} & NL $z=0$               & CNN+HE18 $z=1$ & NL $z=1$              \\ \hline
\multirow{3}{*}{0.1\ $h$/Mpc} & 
10 $\hMpc$        & 52.7 & 69.9                                                   & \multirow{3}{*}{103.2} & 57.2         & \multirow{3}{*}{76.2} \\
&5 $\hMpc$         & 48.0 & 64.7                                                   &                        & 52.4         &                       \\
&Cosine            & 46.4 & 52.9                                                   &                        & 50.7         &                       \\ \hline
0.2\ $h$/Mpc &
Cosine            & 15.8 & -                                                      & -                      & 17.4         & 54.5                  \\ \hline
\end{tabular}
\caption{Fisher forecast for the error in $\fnl$, $\sigma(\fnl)$, from $\langle\Phi^2\delta\rangle$ estimator applied on CNN+HE18 reconstructed fields in comparison with pre-reconstruction (the ``NL'' columns) and the initial condition cases at $z=0$ and 1 and for $k_{\rm max}=0.1$ and 0.2 $h$/Mpc. The ``Smoothing'' column denotes the smoothing scales for the IC and CNN+HE18 cases. The cosine filter is applied for $k$ in $0.2-0.25\ h$/Mpc. The two nonlinear cases are both smoothed at 20 $\hMpc$. The cosine filter cases most closely match the corresponding initial condition forecasts, and the agreement is similar at both redshifts although $z=1$ is better. Comparing to the nonlinear case at $z=1$, we achieve a factor of 1.5 improvement in $\sigma(\fnl)$ estimate with CNN+HE18 reconstruction with $k_{\rm max}=0.1\ h$/Mpc and a factor of 3 with $k_{\rm max}=0.2\ h$/Mpc.   }
\label{tab:forecast}
\end{table}

\section{Information content at the field level}\label{sec:template}

\subsection{Perturbation theory model}\label{sec:theory_PT}

To better understand the discrepancy in the cross-power spectra shown in the previous section, and to 
quantify the impact reconstruction has on the PNG signal, we turn to modeling the reconstructed fields directly.
We start with a summary of perturbation theory, which we will build upon to fit both the nonlinear and 
reconstructed fields. At the second order in Eulerian perturbation theory and in the absence of primordial non-Gaussianity, the nonlinear matter density field can be written as a sum of four density components. 
In configuration space, we have 
\begin{equation}\label{eqn:expansion}
    \delta(\boldsymbol{x})=b_{G}\delta_{G}(\boldsymbol{x})+b_{2}\delta_{G}(\boldsymbol{x})^2+b_{\nabla^2}\delta_{\nabla^2}(\boldsymbol{x})+b_{s^2}\delta_{s^2}(\boldsymbol{x}).
\end{equation}
Here, $\delta_{G}$ is the linear Gaussian density field. 
The growth term, $\delta_{G}^2$,
captures the nonlinear growth of perturbations. 
The shift term, $\delta_{\nabla^2}$, describes the motion of density due to gravity, while the tidal term, $\delta_{s^2}$, describes the anisotropic distortion of the perturbations \cite{Sherwin12}. Specifically, the shift term is defined as
\begin{equation}
    \delta_{\nabla^2}(\boldsymbol{x})=-\boldsymbol{\Psi}(\boldsymbol{x})\cdot \nabla\delta_{G}(\boldsymbol{x}),
\end{equation}
where $\boldsymbol{\Psi}(\boldsymbol{x})$ is the first order displacement. In Fourier space, this can be calculated as $\boldsymbol{\Psi}(\boldsymbol{k})=-i(\boldsymbol{k}/k^2) \delta_{G}(\boldsymbol{k})$. The tidal term is defined as
\begin{equation}
    \delta_{s^2}(\boldsymbol{x})=\frac{3}{2}s_{ij}(\boldsymbol{x})s^{ij}(\boldsymbol{x}),
\end{equation}
where Einstein summation is implied. Here $s_{ij}(\boldsymbol{x})$ is the tidal tensor, which we calculate in Fourier space: 
\begin{equation}
    s_{ij}(\boldsymbol{k})=\left(\frac{k_i k_j}{k^2}-\frac{1}{3}\delta_{K,ij}\right)\delta_{G}(\boldsymbol{k}).
\end{equation}
Here $\delta_{K,ij}$ is the Kronecker delta function. 
In all three cases, we also subtract the mean of each of these terms (we suppress this for brevity).
In Eulerian perturbation theory (in particular, the $F_2$ kernel), the coefficients of the above 
three quadratic terms are
$b_2=17/21$, $b_{\nabla^2}=-1$ and $b_s^2=4/21$\footnote{In some references, the third coefficient is 2/7 instead of 4/21; the difference is because a factor of 3/2 is included in our tidal term. } \cite[e.g.][]{Schmittfull15},
while $b_G = 1$.

We can extend this model to include PNG
\begin{equation}\label{eqn:expansion_w_delta_fnl}
    \delta(\boldsymbol{x})=b_{G}\delta_{G}(\boldsymbol{x})+f_{\rm NL}\delta_{f_{\rm NL}}(\boldsymbol{x})+b_{2}\delta_{G}(\boldsymbol{x})^2+b_{\nabla^2}\delta_{\nabla^2}(\boldsymbol{x})+b_{s^2}\delta_{s^2}(\boldsymbol{x}), 
\end{equation}
where the definition of $\delta_{f_{\rm NL}}$ follows from Eqns.~\ref{eq:local_potential} and~\ref{eq:delta_to_Phi}.
We use the same model for both the nonlinear and the reconstructed fields. 
For the reconstructed field (both CNN+HE18 and HE18), this is an ansatz. 
We expect reconstruction to reduce the amplitude of the quadratic bias terms.

We use the $\fnl=0$ initial condition density to approximate $\delta_G$ for all fitting cases\footnote{Recall
that the $f_{\rm NL}=\pm 100$ simulations started with the same initial seed, and therefore, the same 
Gaussian field as the $f_{\rm NL}=0$ case.}.
The $\delta_{\fnl}$ term is computed using two initial condition density fields with different $f_{\rm NL}$ values, assuming the dependence on $\fnl$ is linear. In our specific case, we have 
\begin{equation}
    \delta_{f_{\rm NL}}=\frac{1}{200}\left[\delta_{\rm IC}^{\fnl=100}-\delta_{\rm IC}^{\fnl=-100}\right].
\end{equation}
Note that using the simulations to define $\delta_G$ and $\delta_{f_{\rm NL}}$ ensures that the phases 
in our model exactly match what we expect from the simulation.
When computing the quadratic terms, we smooth $\delta_G$ by a 3 $h^{-1}$ Mpc Gaussian. We also smooth the linear term 
$\delta_G$ to account for the erasure of initial conditions by higher order perturbations not captured by our expansion, i.e. applying a damping on the linear term $\delta_{G}$.
We treat this as a parameter optimized for different fields (see below).

\subsection{Fitting procedure}\label{sec:fitting_procedure}

A perfect fit would have coefficients $(b_0,f_{\rm NL}, b_2, b_{\nabla^2},b_{s^2})=(1, 0, 17/21, -1, 4/21)$ 
for the nonlinear case 
and $(b_0,f_{\rm NL}, b_2, b_{\nabla^2},b_{s^2})=(1, 0, 0, 0, 0)$ for both the CNN+HE18 and the HE18 
reconstruction cases. This assumes that reconstruction removes 
all second-order nonlinearities. We use this to define the residual power spectrum
\begin{equation}\label{eqn:residuel}
    P_{\rm residual}(k)= \langle |\delta(\boldsymbol{k})-\delta_{\rm perfect}(\boldsymbol{k})|^2\rangle/V.
\end{equation}
In this expression, $\delta(\boldsymbol{k})$ can be the reconstructed or nonlinear density, and $V$ is the volume.

\begin{figure}
    \centering
    \includegraphics[width=0.6\linewidth]{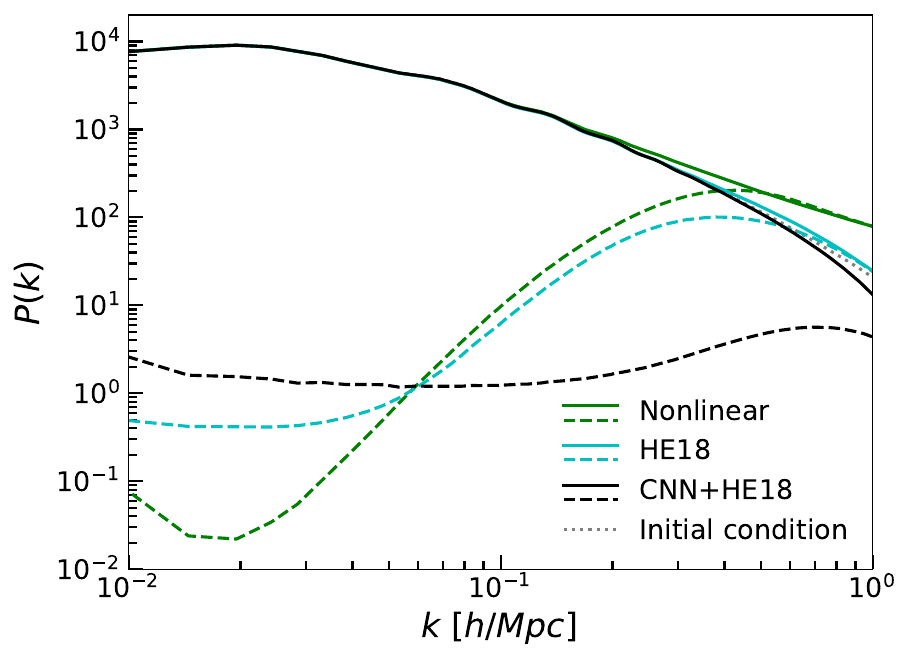}
    \caption{Reconstructed and nonlinear power spectrum (solid) as well as
    residual power spectrum (dashed) for CNN+HE18 (black), HE18 (cyan), and
    nonlinear (green) $\fnl=0$ cases at $z=1$. We also show the initial power
    spectrum as the grey dotted line. The residual power spectrum is defined in
    Eqn.~\ref{eqn:residuel}. 
    The damping scale applied is 3 $\hMpc$ for the nonlinear, 2 $\hMpc$ for the
    HE18, and 1 $\hMpc$ for the CNN+HE18 cases.
    The residual is small for the nonlinear case on large scales, suggesting
    that perturbation theory accurately describes the nonlinear density on these scales.
    We also observe that the hybrid reconstruction field is well described by the linear 
    density field out to $k \sim 0.2 h/$Mpc, with very little scale-dependence seen in the 
    residual.}
    \label{fig:residual}
\end{figure}

Figure~\ref{fig:residual} shows the residual power spectrum for nonlinear as well as reconstruction cases, compared to the initial and nonlinear/reconstructed power spectra at $z=1$. The residual is slightly larger in the reconstruction cases than in the nonlinear cases at large scales. 
This suggests that our model of the nonlinear density field is a better description at these scales 
than the idealized model assuming complete removal of gravitational nonlinearity for the reconstruction cases.
However, reconstruction keeps residual scale-independent longer, especially with the hybrid method, 
where the residual remains constant to $k\sim0.2\ h$/Mpc at $z=1$.

We fit the nonlinear/reconstructed fields using the model in Eqn.~\ref{eqn:expansion_w_delta_fnl}. 
We compute the model in Fourier space, using the actual $\delta_G$ and $\delta_{f_{\rm NL}}$ fields from the 
simulations. We apply a $k$-cut at $k_{\rm max}$ in Fourier space to discard the
high-$k$ modes. This is needed for both nonlinear and reconstruction fits. In
the nonlinear case, the perturbation theory based model breaks down at higher
$k$; in the reconstruction cases, the reconstruction efficacy decreases going to
higher $k$. We weight the modes in the model and data by $1/\sqrt{P_{\rm residual}}$
using the residual power spectrum shown in Figure~\ref{fig:residual}.
We determine the model parameters with a direct least-squares 
optimization of Eqn.~\ref{eqn:expansion_w_delta_fnl}.

\subsection{Fitting results}\label{sec:fitting_results}

In Table~\ref{tab:fits}, we present fits for the reconstructed and nonlinear density fields. The damping scales for $\delta_G$ are 0, 2, 5 $\hMpc$ for the CNN+HE18, HE18 and nonlinear density cases at $z=0$, respectively, and 1, 2, 3 $\hMpc$ at $z=1$, respectively. These damping values were manually chosen such that $b_G$ is close to 1 for the $\fnl=0$ case. 
We apply a $k_{\rm max}=0.1\ h$/Mpc cut as default for all cases and at both $z=0$ and 1. 
Motivated by Figure~\ref{fig:residual}, we also add two higher $k$-cuts for CNN+HE18, with $k_{\rm max}=0.15\ h$/Mpc for $z=0$ and $k_{\rm max}=0.2\ h$/Mpc for $z=1$. 
The error is the standard deviation of fitting 90 simulations. 
For the nonlinear fields, the fits for the quadratic terms are similar to the expected values from 
perturbation theory.
After reconstruction, these quadratic terms are much lower, especially with CNN+HE18, suggesting that reconstruction effectively removes gravity-induced nonlinearity at the second order.

Table~\ref{tab:fits} also shows that reconstruction reduces the error in $\fnl$ compared to the corresponding nonlinear cases, with the hybrid reconstruction giving lower errors. At the same $k$-cut, we get about 1.4$\times$ improvement at $z=0$ with hybrid reconstruction over the nonlinear field, and 1.5$\times$ at $z=1$. 
With a higher $k_{\rm max}$, the error can be even lower. However, going to higher $k_{\rm max}$ with our model is only possible for the CNN+HE18 cases. As shown in Figure~\ref{fig:residual}, the residual for HE18 and nonlinear densities quickly increases above $k=0.1\ h$/Mpc. The inverse variance weighting is effective only for mildly scale-dependent residuals. Extending $k_{\rm max}$ from 0.1 to 0.2 $h$/Mpc increases the residual power spectrum for the nonlinear and HE18 cases by nearly an order of magnitude. In contrast, the CNN+HE18 residual remains scale-independent longer, so it is possible to have a higher $k$-cut for CNN+HE18, which allows us to probe more modes and lower the error in $\fnl$. 
However, the bias in the resulting $f_{\rm NL}$ values becomes larger with a larger $k_{\rm max}$ for both the zero and non-zero $\fnl$ cases.

\begin{figure}
    \centering
    \includegraphics[width=0.6\columnwidth]{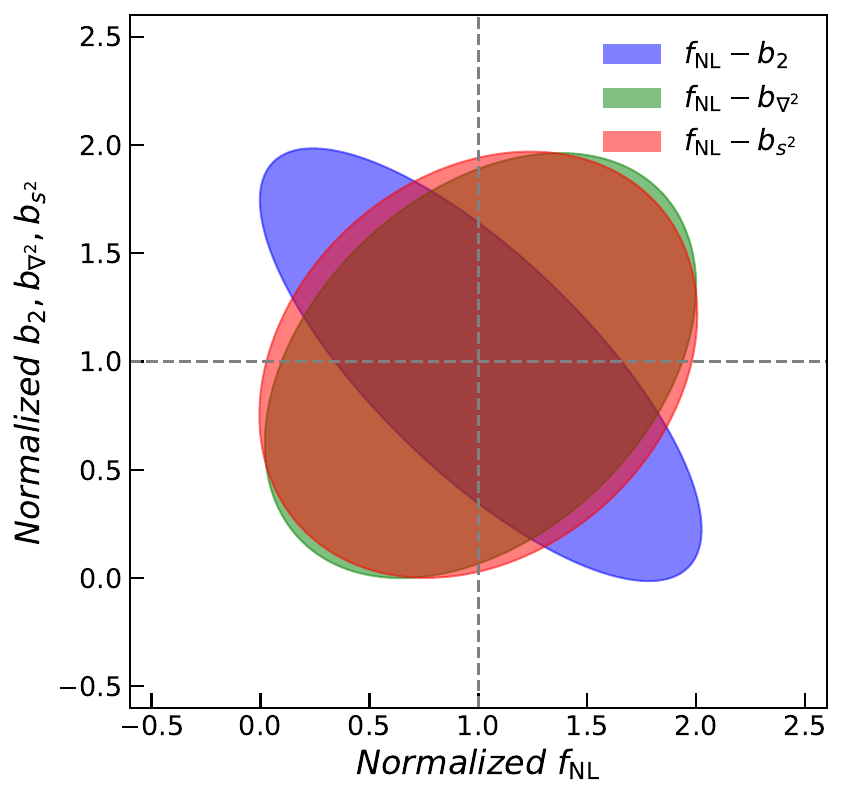}
    \caption{1-$\sigma$ correlation coefficient ellipses between $f_{\rm NL}$ and the three quadratic bias terms, for fitting of the CNN+HE18 reconstructed density for $f_{\rm NL}=0$ at $z=1$. The bias terms as well as $\fnl$ are normalized by their individual means. $f_{\rm NL}$ and $b_2$ have the most degeneracy, since they are the monopole terms. This suggest that the quadratic bias terms need to be constrained together with $f_{\rm NL}$, or the error in $f_{\rm NL}$ will be underestimated. The degeneracy between $\fnl$ and $b_G$ is nearly zero, so we do not include it here. }
    \label{fig:degeneracy}
\end{figure}

Additionally, we find that there are degeneracies between $\fnl$ and the three gravitational bias terms, with $\fnl$ and $b_2$ having the strongest degeneracy. Figure~\ref{fig:degeneracy} shows the correlation coefficient between the fitted $f_{\rm NL}$ and the three gravitational bias parameters for the $z=1$ $\fnl=0$ simulations. This degeneracy suggests that we need to constrain $\fnl$ and the quadratic bias terms together, otherwise we would underestimate the $\fnl$ error. The correlation between $\fnl$ and $b_G$ is nearly zero. Appendix~\ref{appx:b2_fnl} explores the correlation between $b_2$ and $\fnl$ using a Fisher forecast, and finds that simultaneously measuring $b_2$ and $\fnl$ would inflate the $\fnl$ errors
by $\sim 30\%$.

\begin{table}[]
\resizebox{\textwidth}{!}{
\begin{tabular}{cccccc}
\hline
                  & $b_{G}$           & $f_{\rm NL}$  & $b_2$           & $b_{\nabla^2}$   & $b_{s^2}$       \\ \hline
\multicolumn{6}{l}{{$z=0$}}\\

\multicolumn{6}{l}{CNN+HE18}                                                                      \\
$f_{\rm NL}$=0    & 0.9995$\pm$0.0007 & -11.7$\pm$4.4    & 0.0048$\pm$0.0009 & -0.0128$\pm$0.0009 & 0.0101$\pm$0.0010 \\
$f_{\rm NL}$=100  & 1.0011$\pm$0.0007 & 80.7$\pm$4.5 (92.4)    & 0.0043$\pm$0.0009 & -0.0129$\pm$0.0009 & 0.0107$\pm$0.0010 \\
$f_{\rm NL}$=-100 & 0.9980$\pm$0.0007 & -103.7$\pm$4.3 (-92.0)   & 0.0053$\pm$0.0008 & -0.0127$\pm$0.0009 & 0.0095$\pm$0.0009 \\

\multicolumn{6}{l}{CNN+HE18, $k_{\rm max}=0.15\ h$/Mpc}                                                                      \\
$f_{\rm NL}$=0    & 0.9990$\pm$0.0008 & -14.9$\pm$4.2    & 0.0033$\pm$0.0007 & -0.0104$\pm$0.0006 & 0.0063$\pm$0.0007 \\
$f_{\rm NL}$=100  & 1.0009$\pm$0.0008 & 74.5$\pm$4.4 (89.4)    & 0.0028$\pm$0.0007 & -0.0103$\pm$0.0006 & 0.0066$\pm$0.0008 \\
$f_{\rm NL}$=-100 & 0.9971$\pm$0.0008 & -103.8$\pm$4.1 (-88.9)   & 0.0038$\pm$0.0007 & -0.0105$\pm$0.0006 & 0.0060$\pm$0.0007 \\

\multicolumn{6}{l}{HE18}                                                                          \\
$f_{\rm NL}$=0    & 0.9955$\pm$0.0002 & -3.5$\pm$4.5    & 0.065$\pm$0.001 & -0.042$\pm$0.002 & -0.018$\pm$0.001 \\
$f_{\rm NL}$=100  & 0.9954$\pm$0.0002 & 92.9$\pm$4.6 (96.4)   & 0.065$\pm$0.001 & -0.042$\pm$0.002 & -0.018$\pm$0.001 \\
$f_{\rm NL}$=-100 & 0.9955$\pm$0.0002 & -100.1$\pm$4.5 (-96.6) & 0.065$\pm$0.001 & -0.042$\pm$0.002 & -0.018$\pm$0.001 \\ 

\multicolumn{6}{l}{Nonlinear}                                                                          \\
$f_{\rm NL}$=0    & 0.9998$\pm$0.0003 & 4.9$\pm$6.1    & 0.791$\pm$0.004 & -0.985$\pm$0.005 & 0.192$\pm$0.003 \\
$f_{\rm NL}$=100  & 0.9997$\pm$0.0003 & 102.6$\pm$6.0 (97.7)   & 0.790$\pm$0.004 & -0.984$\pm$0.006 & 0.193$\pm$0.003 \\
$f_{\rm NL}$=-100 & 0.9999$\pm$0.0003 & -92.8$\pm$6.4 (-97.7) & 0.791$\pm$0.003 & -0.985$\pm$0.004 & 0.192$\pm$0.002 \\ \hline

\multicolumn{6}{l}{$z=1$}\\
\multicolumn{6}{l}{CNN+HE18}                                                                      \\
$f_{\rm NL}$=0    & 0.9998$\pm$0.0006 & 0.6$\pm$3.1    & 0.004$\pm$0.001 & -0.018$\pm$0.001 & 0.017$\pm$0.001 \\
$f_{\rm NL}$=100  & 1.0015$\pm$0.0006 & 93.0$\pm$3.2 (92.4)    & 0.004$\pm$0.001 & -0.018$\pm$0.001 & 0.018$\pm$0.001 \\
$f_{\rm NL}$=-100 & 0.9980$\pm$0.0006 & -91.7$\pm$3.1 (-92.3)   & 0.004$\pm$0.001 & -0.018$\pm$0.001 & 0.016$\pm$0.001 \\

\multicolumn{6}{l}{CNN+HE18, $k_{\rm max}=0.2\ h$/Mpc}                                                                      \\
$f_{\rm NL}$=0    & 0.9970$\pm$0.0006 & -8.5$\pm$2.4    & 0.0042$\pm$0.0006 & -0.0154$\pm$0.0005 & 0.0182$\pm$0.0006 \\
$f_{\rm NL}$=100  & 0.9988$\pm$0.0006 & 82.4$\pm$2.4 (90.9)    & 0.0040$\pm$0.0006 & -0.0155$\pm$0.0005 & 0.0190$\pm$0.0006 \\
$f_{\rm NL}$=-100 & 0.9951$\pm$0.0006 & -99.3$\pm$2.3 (-90.8)   & 0.0044$\pm$0.0006 & -0.0153$\pm$0.0005 & 0.0174$\pm$0.0006 \\

\multicolumn{6}{l}{HE18}                                                                          \\
$f_{\rm NL}$=0    & 1.0005$\pm$0.0002 & 7.4$\pm$3.5    & 0.076$\pm$0.002 & -0.061$\pm$0.002 & -0.008$\pm$0.002 \\
$f_{\rm NL}$=100  & 1.0006$\pm$0.0002 & 106.0$\pm$3.5 (98.6)   & 0.076$\pm$0.002 & -0.061$\pm$0.002 & -0.008$\pm$0.002 \\
$f_{\rm NL}$=-100 & 1.0005$\pm$0.0002 & -91.2$\pm$3.5 (-98.6) & 0.076$\pm$0.001 & -0.061$\pm$0.002 & -0.007$\pm$0.002 \\ 

\multicolumn{6}{l}{Nonlinear}                                                                          \\
$f_{\rm NL}$=0    & 0.9994$\pm$0.0002 & -5.8$\pm$4.5    & 0.820$\pm$0.003 & -1.023$\pm$0.004 & 0.202$\pm$0.002 \\
$f_{\rm NL}$=100  & 0.9994$\pm$0.0002 & 92.6$\pm$4.3 (98.4)   & 0.820$\pm$0.004 & -1.023$\pm$0.005 & 0.202$\pm$0.002 \\
$f_{\rm NL}$=-100 & 0.9994$\pm$0.0002 & -104.4$\pm$4.7 (-98.6) & 0.820$\pm$0.003 & -1.023$\pm$0.003 & 0.201$\pm$0.002 \\ \hline

\end{tabular}
}

\caption{
Results of fitting templates in
the form of Eqn.~\ref{eqn:expansion_w_delta_fnl} (see text for details of 
the fitting procedure). The errors are the standard deviation of the 90 simulations, 
and not the error on the mean values quoted.
The values in the parentheses are the mean $\fnl$ values after removing the
shift at $\fnl=0$.
The quadratic terms are largely reduced after
reconstruction, especially for the CNN+HE18 cases.
Reconstruction reduces the error in the $\fnl$ values at comparable $k$ values, 
and allows the fit to access larger $k$'s to reduce the error further. 
However, we observe bias in both
zero and non-zero $\fnl$ cases. 
}
\label{tab:fits}
\end{table}

\subsubsection{Modeling the cross-power}

\begin{figure}
    \centering
    \includegraphics[width=0.6\linewidth]{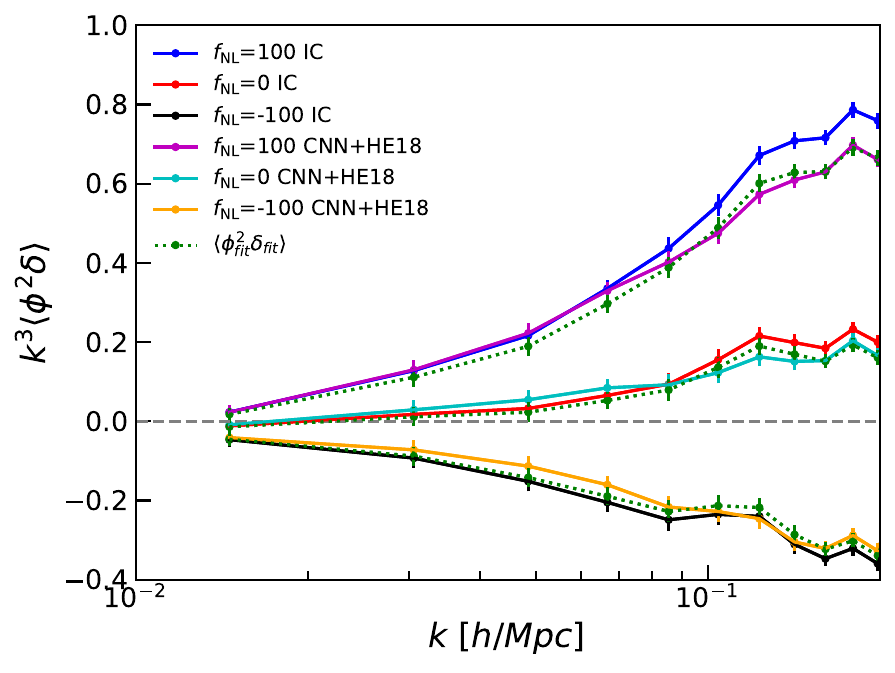}
    \caption{Cross-power using the fit to the reconstructed density $\delta_{\rm rec}$ up to $k_{\rm max}=0.2\ h$/Mpc for the cosine filter case (green dotted line for each $\fnl$ value). All other lines are the same as the right panel of Figure~\ref{fig:statistic}. The cross-power with the fitted reconstructed density is consistent with the measured reconstructed case within the error bars, suggesting that we can model the reconstructed cross-power with our perturbative model.  }
    \label{fig:cross-power_fit_density}
\end{figure}

We use this model to predict our cross-power estimator.
We use the template fit results to approximate the reconstructed density, and this fitted density is then used to compute $\Phi_{\rm rec}^2$. We then cross-correlate the two to get the fitted cross-power.
Figure~\ref{fig:cross-power_fit_density} shows the result of the fitted cross-power in comparison to the cross-power of the reconstruction and initial condition versions for the cosine filter case (the right panel of Figure~\ref{fig:statistic}). We use the fitting results for $k_{\rm max}=0.2\ h$/Mpc as shown in Table~\ref{tab:fits}. The fitted result is consistent with the reconstructed cross-power within error bars for all three $\fnl$ values. 
The agreement here indicates that we can model the reconstructed cross-power 
with a simple
perturbative model. This is encouraging for the application of the method developed in this work to real surveys.

% \subsubsection{Discussion of the fitting results}
\subsubsection{Bias in $\fnl$}
\label{sec:discussion_fitting_results}
As Figure~\ref{fig:cross-power_fit_density} shows, much of the discrepancy between the cross-power spectrum applied to the reconstructed and IC fields can be accounted for with a simple perturbative model. However, as Table~\ref{tab:fits} shows, the measured  $\fnl$ values within this perturbative model are biased from their true values. 

Our simulation results suggest two kinds of $\fnl$ biases. The first is an additive bias that we see in the $\fnl=0$ simulations, at a level of $\Delta \fnl = 0.6 \pm 0.3$ at $z=1$ and $\Delta \fnl = -11.7 \pm 0.5$ at $z=0$. Interestingly, we also measure a bias for the nonlinear simulations of $\Delta \fnl=-5.8 \pm 0.5$ at $z=1$ and $\Delta \fnl = 4.9 \pm 0.6$ at $z=0$.
Removing this shift from the $\fnl = \pm 100$ simulations results in symmetric $\fnl$ measurements for these cases (adjusted mean presented in the parentheses in the table), further strengthening the suggestion that this is an additive effect. The change of this bias with decreasing redshift, the simultaneous presence of bias in the fits of the quadratic bias terms, and the fact that we observe it for both reconstructed and non-reconstructed simulations suggests that this might be caused by imperfectly modeled nonlinearities and/or residual artifacts in our fitting procedure. We also find this bias decreases with a smaller $k_{\rm max}$, further supporting this hypothesis.

Accounting for this additive shift, we find a residual shift of $\Delta \fnl \sim \pm 10$ for the $\fnl = \pm 100$ simulations. Unlike the additive shift, this shift is much smaller for the nonlinear simulations, and even for the HE18 case, suggesting that the hybrid reconstruction is distorting the PNG signal. Since the hybrid reconstruction was trained on $\fnl=0$ simulations and the shift reduces the $\fnl$ signal, one might guess that the neural network was over-Gaussianizing the model. To test this, we rerun the training using $\fnl = \pm 100$ simulations in addition to the $\fnl = 0$ simulations. However, we found no improvements in the performance of the network.

Another possible source of bias is the effect of $\fnl$ on the nonlinear terms. In particular, the 
quadratic terms in our model depend on the linear density field (including the $\fnl$ term), not just the Gaussian 
field. We test for this in two ways - first by using the full linear density field to compute the quadratic
bias terms, and second, by introducing a $\delta_G \delta_{\fnl}$ term to our model. Neither of these 
reduced the bias in the $\fnl$ values.

Understanding and calibrating these biases will require simulations with a range of $\fnl$ values. It is possible that these biases can simply be calibrated using simulations. A better understanding of how the network affects the PNG signal may also suggest modifications to the architecture to better preserve the $\fnl$ signal.

\subsection{Robustness to change of cosmology}

The above assumed that we had the correct cosmology (and the initial conditions) when computing our model. 
To test the robustness of our conclusions to an incorrect cosmology, we consider a set of simulations with 
$\Omega_{\rm m}$ varied by $\pm 3\%$ and with $\fnl=0$. We analyze these identically to the fiducial simulations.
In particular, we assume the fiducial $\Omega_{\rm m}$ values in our code, and do not retrain the CNN for 
the hybrid reconstruction. Furthermore, we use the fiducial $\delta_G$ and $\delta_{\fnl}$ fields when computing 
our model, using the fact that these new simulations were phase-matched to the fiducial cosmology (similar to the
$\fnl$ simulations). 

Table~\ref{tab:Om} shows the fitting results for these simulations. While we do see shifts in 
the recovered $\fnl$ values, these shifts are smaller than the biases for the $\fnl=0$ cases discussed above. 
The changes in the $b_G$ values are likely due to differences in the shape of the power spectrum. 
Interestingly, we find that the quadratic bias terms, which describe the gravitational evolution, are rather stable with the change of $\Omega_{\rm m}$ after the hybrid reconstruction, which suggests that CNN reconstruction is generalizable to changes in $\Omega_{\rm m}$ (we already demonstrate the generalizablity of our reconstruction to other cosmologies in \cite{CNN}). By contrast, for both the nonlinear and HE18 fits, 
the quadratic bias terms change with $\Omega_{\rm m}$. These effects will need to be accounted for when applying 
this method to data.

\begin{table}[]
\begin{tabular}{cccccc}
\hline
                  & $b_{G}$           & $f_{\rm NL}$  & $b_2$           & $b_{\nabla^2}$   & $b_{s^2}$       \\ \hline
\multicolumn{6}{l}{{$z=1$}}\\

\multicolumn{6}{l}{CNN+HE18}                                                                      \\
$\Omega_{\rm m}=0.3075$   & 1.0159$\pm$0.0006 & 0.9$\pm$3.4    & 0.004$\pm$0.001 & -0.018$\pm$0.001 & 0.018$\pm$0.001 \\
$\Omega_{\rm m}=0.3275$  & 0.9842$\pm$0.0006 & 0.3$\pm$3.3  & 0.004$\pm$0.001 & -0.017$\pm$0.001 & 0.016$\pm$0.001 \\

\multicolumn{6}{l}{HE18}                                                                      \\
$\Omega_{\rm m}=0.3075$   & 1.0219$\pm$0.0003 & 9.7$\pm$4.0    & 0.078$\pm$0.002 & -0.063$\pm$0.002 & -0.007$\pm$0.002 \\
$\Omega_{\rm m}=0.3275$  & 0.9801$\pm$0.0003 & 5.3$\pm$4.5  & 0.074$\pm$0.002 & -0.058$\pm$0.002 & -0.008$\pm$0.002 \\

\multicolumn{6}{l}{Nonlinear}                                                                      \\
$\Omega_{\rm m}=0.3075$    & 1.0291$\pm$0.0004 & 6.4$\pm$5.3    & 0.860$\pm$0.002 & -1.080$\pm$0.004 & 0.219$\pm$0.003 \\
$\Omega_{\rm m}=0.3275$  & 0.9714$\pm$0.0005 & -16.3$\pm$7.5    & 0.783$\pm$0.005 & -0.971$\pm$0.005 & 0.186$\pm$0.003 \\
\hline

\end{tabular}
\caption{Template fits for simulations with $\fnl=0$ at $z=1$ with $\Omega_{\rm m}$ varied by 3\%. Both reconstruction routines are done assuming the fiducial $\Omega_{\rm m}$, so this is testing the effect of a deviation from fiducial cosmology on the $\fnl$ constraints. Comparison to fits for the fiducial cases presented in Table~\ref{tab:fits}, we find that $\fnl$ responds to changes in $\Omega_{\rm m}$, but the changes are smaller than the bias at $\fnl=0$. 
}
\label{tab:Om}
\end{table}

\section{Discussion}\label{sec:discussion}

We start by discussing how to address the complications of applying this work to real surveys. We then discuss 
applications of our method to other forms of PNG. Finally, we compare with prior work.

\subsection{Complications with applications to real surveys}

In this work, we only focus on real space matter density field. To apply this
method to real surveys, we need to consider shot noise, galaxy bias, redshift
space distortions (RSD), survey geometry, and other complications associated with
real surveys. We comment on the first four below in turn.

At the number density used in this paper, $\bar{n} \sim10^{-1}\ (h/{\rm
Mpc})^3$, the reconstructed field (by our hybrid reconstruction) is 90\% correlated with the 
initial density field up to $k\sim0.48\ h$/Mpc at
$z=0$. At number densities more characteristic of current galaxy surveys,
$\bar{n} \sim 10^{-4}-10^{-3}\ (h/{\rm
Mpc})^3$ (e.g. DESI luminous red galaxy (LRG) and emission line galaxy (ELG)
samples, ignoring the effect of galaxy bias), 
this scale drops to 
$k\sim 0.13\ h$/Mpc at $z=0$. The quasar samples previously
used for $\fnl$ measurements have even lower number densities 
and present an even bigger challenge. More optimistically, lower 
redshift samples like the DESI Bright Galaxy Survey (BGS) \cite{Hahn23}
can reach number densities of $\bar{n} \sim 10^{-2}\ (h/\rm Mpc)^3$. Furthermore,
reconstruction has the potential to extend the $k_{\rm max}$ for these 
analyses, mitigating the effect of the smaller volumes. 
In all of these 
cases, the question is whether reconstruction has the ability to 
reduce the gravitational bispectrum signal to enhance the underlying 
primordial signal, and if the residual imperfections in the reconstruction can be 
modeled better than the full nonlinear evolution of the density field. 
This is still an open question, but the results presented here motivate 
exploring this in more detail. The potential gains that a higher number density
might offer could also inform design decisions for future surveys.
Finally, as noted in Section~\ref{sec:derive_cross_power}, the cross-power expression needs to be modified in the presence of shot noise. Extra parameters need to be introduced to fit shot noise, but the overall structure of the cross-power stays the same.

Galaxy bias shows up conceptually in two locations. The first is that the number 
of parameters that one would need to simultaneously measure increases due to the 
different galaxy bias parameters that show up at the level of the bispectrum. 
Degeneracies between these parameters and $\fnl$ may degrade the $\fnl$
signal. The second place that galaxy bias appears is in the reconstruction 
step itself. In particular, for hybrid reconstruction techniques, the best 
way to train the network in the presence of galaxy bias is still being actively
investigated. Finally, it is worth noting that a biased galaxy sample can 
boost a PNG signal (as in the case of the local type models), partially
balancing other losses.

The hybrid reconstruction algorithm presented here is effective 
at removing RSD. As shown in \cite{CNN},
the reconstructed quadrupole in redshift space is nearly zero on all scales,
including Finger-of-God effects on small scales.
We can use the same reconstruction routine for
fields in redshift space to reconstruct the initial condition in real
space. 
While this will need to be calibrated with simulations, we expect a 
very similar approach to that presented here to be applicable. 

The survey geometry can affect the hybrid reconstruction through the 
presence of boundaries. This is well studied in the context of reconstructing 
the BAO signal. Extending it to the hybrid approach could involve incorporating 
boundaries into the training data or modifying the architecture of the network 
to account for the missing data.

\subsection{Applications beyond the local $\fnl$}

In galaxy surveys, unlike the local type, the non-local PNG do not produce a sizeable scale-dependent bias on the galaxy power spectrum. Thus, the measurement of the non-local PNG must rely on the bispectrum. 
Attempts to measure the non-local types of PNG with LSS surveys are still in their infancy and yield errors that are a factor of a few worse than the CMB constraints. However, these measurements will be improved with future experiments.
The method we introduce in this paper can potentially be more helpful for the non-local types, because these cannot be measured with power spectrum alone.

Our method can be extended to other types of PNG that have analytical forms. The general cross-power between a quadratic field and a linear field requires the bispectrum to be product-separable. Since both equilateral and orthogonal shapes are product-separable, we can develop similar cross-power estimators for both these two shapes.
The non-local type primordial potential contains a
kernel in the convolution of the primordial potential \cite[e.g.][]{Coulton23}.
An interesting question for future work is to explore simple yet optimal
estimators for these non-local types of PNG.

For the equilateral shape, reconstruction could be more useful, although the requirement for reconstruction might also be higher. This is because gravity induced nonlinearity couples modes of about the same length, therefore the gravity induced bispectrum peaks at roughly the equilateral shape. The contamination from nonlinear evolution for the primordial equilateral shape is therefore higher compared to other types of PNG. For the local type, the squeezed limit bispectrum would be caused by the local type PNG only, because gravity does not produce this shape of bispectrum. 
This would suggest that reconstruction could further enhance the equilateral PNG signal, relative to the improvements seen in this work.

\subsection{Comparing with previous work}
In this section, we compare our work with two relevant previous studies by Shirasaki et al. \cite{Sugiyama21} and Fl{\"o}ss \& Meerburg \cite{Floss23}. The study of \cite{Sugiyama21} uses the standard reconstruction algorithm \cite{Eisenstein07} and computes the bispectrum using a decomposition formalism by \cite{Suyigama19}. They use simulations of halo fields with a number density of 3.9$\times 10^{-4}(\hMpc)^{-3}$ and linear bias $b=1.8$ at $z=0.484$ in redshift space in a 4 $(h^{-1}$Gpc$)^3$ volume. They derive a model for the lowest-order monopole bispectrum and a leading order anisotropic bispectrum using standard perturbation theory at 1-loop order, and forecast for $\fnl$ together with $b_2$ and $b_{s^2}$ (note that these are halo bias parameters and are different from our $b_2$ and $b_{s^2}$). With these two bispectra and $k_{\rm max}=0.2\ h$/Mpc, they forecast a marginalized error of $\sigma(\fnl)=13.3$ and an unmarginalized error of $\sigma(\fnl)=9.1$ post-reconstruction, which is a factor of 3.2 or 4.2 improvement from pre-reconstruction. Compared to this result, our measurement error $\sigma(\fnl)\sim50$ with reconstruction in 1 $(h^{-1}$Gpc$)^3$ volume at either redshift with $k_{\rm max}=0.1\ h$/Mpc using $\langle\Phi^2\delta\rangle$ implies that with $4\ (h^{-1}$Gpc$)^3$ survey volume, we forecast a measurement error of $\sigma(\fnl)\sim25$. With $k_{\rm max}=0.2\ h$/Mpc, we forecast $\sigma(\fnl)\sim 9$ in $4\ (h^{-1}$Gpc$)^3$ at $z=1$, which is comparable to their result (although at a higher redshift). We also only have a factor of about 2 improvement at $z=0$ (with $k_{\rm max}=0.1\ h$/Mpc) and a factor of 3 improvement at $z=1$ in constraining power after reconstruction with our cross-power estimator. 
Moreover, there is a discrepancy between their data and theory, shown in their Figure 1, 
which could bias their estimates as well; their paper does not address this possible bias.

The study of \cite{Floss23} uses a U-Net for reconstruction for the same $\Quijote$-PNG matter field simulations that we use. They use the mean of simulations for the bispectrum as the model and forecast for the error of $\fnl$ together with cosmological parameters. With $k_{\rm max}=0.52\ h$/Mpc, they forecast a marginalized error $\sigma(\fnl)= 18.35$ and an unmarginalized error $\sigma(\fnl)=16.86$ at $z=0$. 
The biggest difference between their results and ours is in the larger $k_{\rm max}$, which is made possible
by their use of simulations to define their theoretical estimates. We have chosen 
a more conservative approach, developing an analytical model for the reconstructed field. 
However, Figures.~\ref{fig:sigma_fnl_kmax} and~\ref{fig:sigma_fnl} both demonstrate that we 
achieve similar constraints for comparable $k_{\rm max}$. It is possible that we can 
robustly extend our model's validity out to smaller scales using a combination of theory and
simulations.

This work 
uses a higher-fidelity reconstruction technique \cite{CNN} compared to these previous work,
as well as a computationally inexpensive estimator of $\fnl$. 
Despite these differences,  we agree on the overarching result
that using the reconstructed density field can improve constraints on PNG. 

The two studies discussed above are the most directly comparable to the results presented in this paper. There have been additional approaches developing new estimators to measure primordial non-Gaussianities. These range from measurements of skew spectra (\cite[][]{Moradinezhad19}, see also \cite{Schmittfull21,Hou24} for development of skew spectra not focusing on PNG, but in redshift space and application for cosmological constraints with simulation-based inference), to producing new auxiliary fields \cite{Kvasiuk24}, to approaches using field-level inference (\cite{Andrews23,Sullivan24,Andrews24}, see also \cite{Nguyen24,Krause24} for field-level inference development not focusing on PNG). Of these, the most direct comparison is with the work on skew spectra, since our estimator has a similar structure. In particular, \cite{Moradinezhad19} also demonstrated that the skew spectra and bispectra have the same information content (similar to our results in Section~\ref{sec:optimality}). Our work diverges from these in our use of the reconstructed density field, and a perturbation theory based modeling of the results (as opposed to simulation-based inference). A detailed comparison of these different approaches is beyond our scope here, but will be important to understand possible systematic errors.

\section{Conclusion}\label{sec:conclusion}

We introduce a new approach to constraining the primordial non-Gaussianity by computing a cross-power estimator $\langle\Phi^2\delta\rangle$ on the reconstructed density and potential fields. 
Our main findings are:
\begin{itemize}
    \item The cross-power $\langle\Phi^2\delta\rangle$ is an optimal estimator of the bispectrum to extract $\fnl$ information. 
    When given access to the same $k$-range, the cross-power and the bispectrum return the same $\sigma(\fnl)$.
    The advantage of cross-power is that it is computationally inexpensive.
    \item Reconstruction, especially with machine learning enhanced version, shows promising constraining power for PNG. It removes much of the gravity induced nonlinearity at the second order and preserves most of the PNG information at the same time. By removing gravitational nonlinearity, reconstruction strengths the signal-to-noise of PNG.
    \item 
    We apply the cross-power estimator $\langle\Phi^2\delta\rangle$ on reconstructed density and potential fields and achieve a factor of 3 improvement over pre-reconstruction in the constraining power for $\fnl$ at $z=1$ with $k_{\rm max}=0.2\ h$/Mpc with single parameter forecast. The improvement factor is dependent on reconstruction performance, which is better at higher redshift. 
    \item Hybrid reconstruction extends the $k-$range over which we can use our
    estimator.
    With higher $k_{\rm max}$, we can probe more $k$-modes and extract more PNG information. We demonstrate in this work that there is benefit of constraining PNG by going to higher $k$, which is a different strategy than the scale-dependent bias approach, which relies on lowest possible $k$-modes and necessitates larger survey volume. 
    \item A simple perturbative model appears to accurately describe our estimator, 
    although we find residual additive and multiplicative biases in the $\fnl$ measurements.
    We speculate that they arise from imperfections in the modeling as well as the 
    reconstruction method distorting the PNG signal. We plan to model and calibrate these 
    biases in future work.
\end{itemize}

While we only analyze the matter density field in real space and for the local type PNG, the promising results found in this study encourages further exploration of using reconstructed products to constrain PNG. There is still work to be done before applying our method to real survey data. We will constrain gravitational bias terms with $\fnl$ simultaneously with the full formalism of the cross-power analysis presented in Section~\ref{sec:multiple_parameters}. We plan to extend our method to realistic galaxy survey data in our subsequent work. 
Finally, we aim to better understand the source of the biases described above and calibrate 
these.
We believe that the method developed in this work provides a computationally inexpensive 
and robust approach to constraining PNG with the current and next generations of LSS surveys.

\appendix

\section{Gaussian covariance for the bispectrum}\label{appx:bk_cov_short}
The covariance matrix for the bispectrum is defined as 
\begin{equation}
\begin{split}
    \boldsymbol{C}^B&={\rm Cov}[\hat{B}(k_1,k_2,k_3),\hat{B}(k'_1,k'_2,k'_3)]\\
    &=\langle\hat{B}(k_1,k_2,k_3)\hat{B}(k'_1,k'_2,k'_3)\rangle-\langle\hat{B}(k_1,k_2,k_3)\rangle\langle\hat{B}(k'_1,k'_2,k'_3)\rangle.
    \end{split}
\end{equation}
We compute this covariance matrix largely following Refs.~\cite{Baldaufnotes} and \cite{Chan17}. We substitute in the above an estimator of the matter bispectrum
\begin{equation}
    \hat{B}(k_1,k_2,k_3)=\frac{1}{N_{\rm triangles}(k_1,k_2,k_3)VV_f^2}\int_{k_1}{\de \boldsymbol{k}_1}\int_{k_2}{\de \boldsymbol{k}_2}\delta(\boldsymbol{k}_1)\delta(\boldsymbol{k}_2)\delta(-\boldsymbol{k}_1-\boldsymbol{k}_2),
\end{equation}
where $N_{\rm trianlges}$ is the number of triangles in a volume consisted of three vectors that make a triangle. This volume can be computed as $V_{123}=\int_{k_1}\de \boldsymbol{k}_1 \de \boldsymbol{k}_2 \de \boldsymbol{k}_3 (2\pi)^3 \delta_D(\boldsymbol{k}_1+\boldsymbol{k}_2+\boldsymbol{k}_3)$, and the integral is between $[k_1-\Delta k/2,k_1+\Delta k/2]$, $[k_2-\Delta k/2,k_2+\Delta k/2]$ and $[k_3-\Delta k/2,k_3+\Delta k/2]$. The volume takes the form of \cite{Chan17} $V_{123}(k_1,k_2,k_3)=8\pi^2k_1k_2k_3(\Delta k)^3\beta(\Delta_{123})$. In this expression, $\Delta_{123}$ is cosine of $\pi$ minus angle between $\boldsymbol{k}_1$ and $\boldsymbol{k}_2$, i.e. 
\begin{equation}
\Delta_{123}=\hat{\boldsymbol{k}_1}\cdot\hat{\boldsymbol{k}_2}=\frac{k_1^2+k_2^2-k_3^2}{2k_1k_2}, 
\end{equation}
which, when taking the side length ordering to be $k_1\geq k_2\geq k_3$, takes value between (0,1]. With this, $\beta(\Delta_{123})=1/2$ for the folded cases ($\Delta_{123}=1$) and $\beta(\Delta_{123})=1$ for all other regular triangles\footnote{These conditions are different from Ref.~\cite{Chan17}, which miscounts the folded cases.  }. Hence, the number of triangles is $N_{\rm triangles}(k_1,k_2,k_3)=V_{123}(k_1,k_2,k_3)/V_f^2$, and $V_f=k_f^3$ is the volume of the fundamental cell where $k_f$ is the fundamental frequency. In the expression of the bispectrum estimator, dividing by $V_f^2$ accounts for the integral units, and $V$ is the volume being considered. Noting that $\langle\hat{B}(k_1,k_2,k_3)\rangle=0$ at the leading order, we can get the diagonal Gaussian covariance for the bispectrum in the form of
\begin{equation}\label{eq:cov_B}
    \boldsymbol{C}^B(k_1,k_2,k_3)=\frac{V}{N_{\rm triangles}(k_1,k_2,k_3)}s_{123}P_{\rm lin}(k_1)P_{\rm lin}(k_2)P_{\rm lin}(k_3).
\end{equation}
Here $s_{123}$ accounts for repetition of different triangle configurations: $s_{123}=6$ for equilateral, $s_{123}=2$ for isosceles, and $s_{123}=1$ for other triangles.

\section{Effect of $b_2$ on estimating $\fnl$}\label{appx:b2_fnl}
While not a focus of this current paper, we show here the impact of omitting $b_2$ in estimating $\fnl$ with the bispectrum. We again use the simplest setup as in Section~\ref{sec:optimality}, where the field is linear, and we calculate the Fisher error analytically. The Fisher matrix is now extended to $2\times 2$ with the $\fnl$ and $b_2$ components on the diagonal and the mixed term on the off-diagonal. Similar to how the $\sigma(\fnl)$ is calculated for the single parameter $\fnl$ case in Section~\ref{sec:optimality}, the Fisher component for $b_2$ uses the theory bispectrum for $b_2$ in the derivative, and the Fisher component for the mixed term uses one derivative for $b_2$ and the other for $\fnl$. The covariance matrix remains the same, because the covariance does not have any assumptions whether the field contains $\fnl$ or $b_2$ and it reflects the Gaussian component of the field.

\begin{figure}[h!]
    \centering
\includegraphics[width=0.6\linewidth]{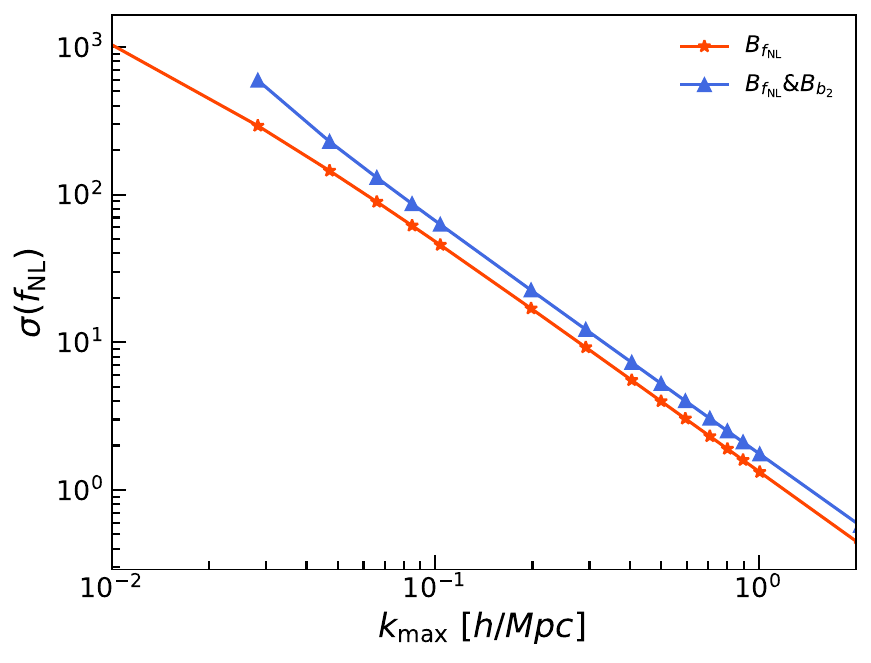}
    \caption{Fisher error $\sigma(\fnl)$ from the bispectrum when both $\fnl$ and $b_2$ are considered (blue), in comparison of $\sigma(\fnl)$ when only $\fnl$ is considered (red, same line as in Figure~\ref{fig:sigma_fnl_kmax}). When both parameters are included, the $\fnl$ error increases. At $k=0.1\ h$/Mpc, the increases is about 30\%. }
    \label{fig:fnl_b2}
\end{figure}

\begin{figure}
    \centering
    \includegraphics[width=0.6\linewidth]{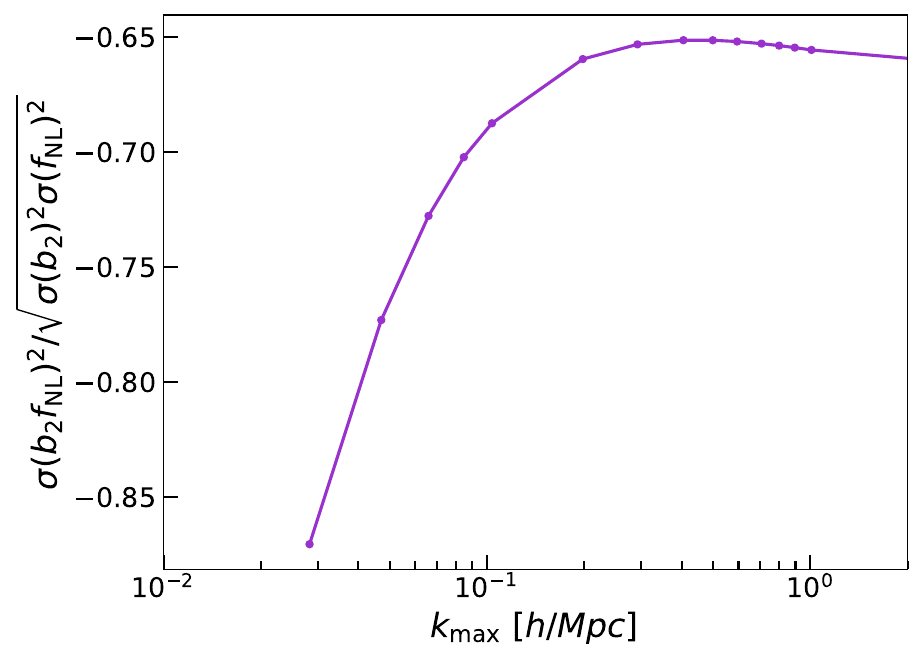}
    \caption{Cross-correlation between $\sigma(b_2)$ and $\sigma(\fnl)$. The cross-correlation between the two decreases as $k_{\rm max}$ becomes higher, which explains why $b_2$ and $\fnl$ does not become more degenerate on smaller scales.}
    \label{fig:sigma_b2_fnl_cross_correlation}
\end{figure}

Figure~\ref{fig:fnl_b2} shows the $\sigma(\fnl)$ from the bispectrum when $b_2$ and $\fnl$ are both considered, in comparison to $\sigma(\fnl)$ when only $\fnl$ is considered. When both are included in the analysis, the $\sigma(\fnl)$ increases as expected. At $k=0.1\ h$/Mpc, the increase is about 30\%. However, as $k$ increases, the impact of $b_2$ becomes smaller. This is likely due to the fact that $\fnl$ and $b_2$ peak at different shapes of triangles. Even though they probe similar fields, i.e. the square of the potential and the square of the density, $\fnl$ is associated with the primordial power spectrum, while $b_2$ the linear matter power spectrum. These two power spectra have different shapes.
Figure~\ref{fig:sigma_b2_fnl_cross_correlation} shows that towards smaller scales, the cross-correlation between $\sigma(\fnl)$ and $\sigma(b_2)$ becomes smaller. As we include more triangles, $\fnl$ and $b_2$ become less degenerate. This suggests that using higher $k$ to constrain $\fnl$, as we suggest in this paper, will not risk more degenerate measurement between $\fnl$ and $b_2$.

\section{Effect of 2LPT in initial condition}\label{appx:2lpt}
\begin{figure}[h]
    \centering
    \includegraphics[width=0.6\columnwidth]{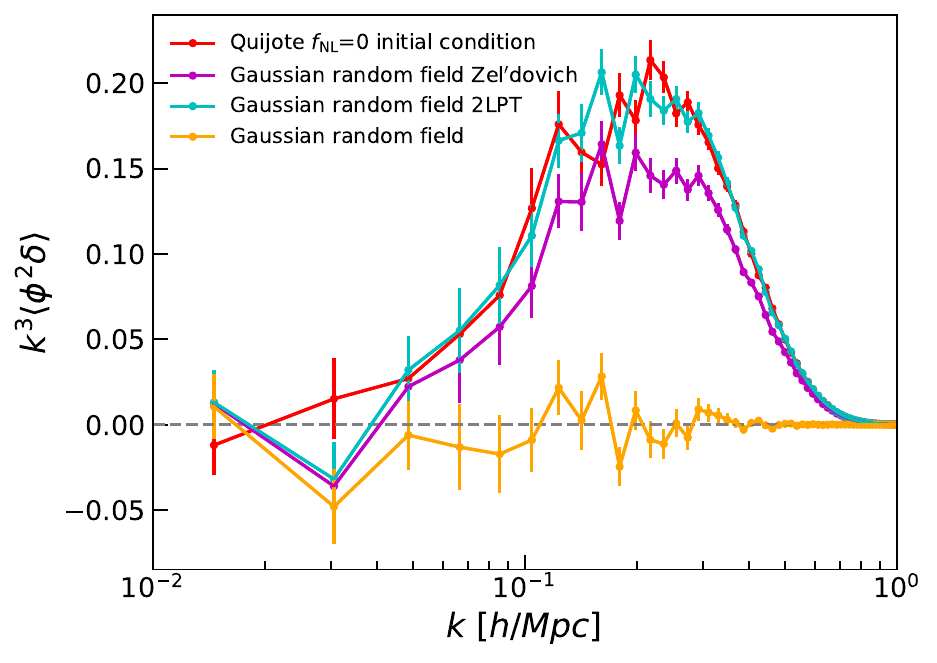}
    \caption{Cross-power of Gaussian random fields with Zel'dovich (magenta), 2LPT (cyan) and by itself (orange), in comparison of cross-power of $\Quijote$ $\fnl=0$ initial condition (red). The cross-power of Gaussian random fields are the average of 100 fields. Here all cases have 5 $\hMpc$ Gaussian smoothing on $\Phi^2$. The red line is same red line in the middle panel of Figure~\ref{fig:statistic}. The Gaussian random field with Zel'dovich contributes most of the observed power at $\fnl=0$, and the 2LPT case matches the $\Quijote$ initial condition.  }
    \label{fig:GRF_IC}
\end{figure}

We show that the fact that when $\fnl=0$ the cross-power is not zero is due to 2LPT. We generate 100 Gaussian random fields that have the linear power spectrum at the $\Quijote$ initial condition redshift $z=127$ on a $512^3$ grid. After that, we compute the Zel'dovich as well as 2LPT displacements, and move particles by these displacements and redistribute particles following the TSC scheme. We calculate the cross-power for these three fields with the application of 5 $\hMpc$ Gaussian smoothing on $\Phi^2$ and compare to the cross-power of $\Quijote$ initial condition with the same smoothing. Figure~\ref{fig:GRF_IC} shows that the cross-power has most contribution from Zel'dovich. When 2LPT is included, the predicted cross-power from the Gaussian random fields agree well with the $\fnl=0$ initial condition of $\Quijote$. Gaussian random fields themselves give cross-power consistent with zero. 

\begin{figure}
    \centering
    \includegraphics[width=0.6\columnwidth]{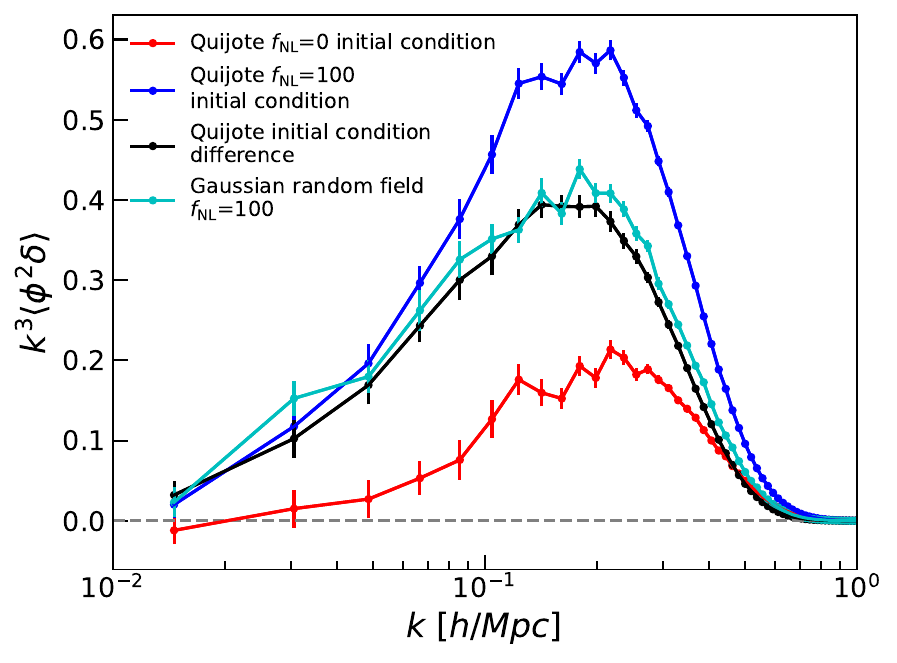}
    \caption{Cross-power of $\fnl=0$ (red) and $\fnl=100$ (blue) initial condition of $\Quijote$ (these match the red and blue lines in Figure~\ref{fig:statistic} middle panel) and the difference between the two cross-powers (black), in comparison with $\fnl=100$ field generated with Gaussian random fields (cyan). The difference between cyan and black is the effect of $\fnl=100$ on the 2LPT field. This figure shows that this effect is small. }
    \label{fig:GRF_100}
\end{figure}

When nonzero $\fnl$ is present in the field, it will have an effect on the nonlinear evolution. We show this effect in Figure~\ref{fig:GRF_100}, where we compare cross-power of $\Quijote$ $\fnl=100$ initial condition subtracted by the cross-power of $\fnl=0$ initial condition with the cross-power of $\fnl=100$ without nonlinear evolution. The $\fnl=100$ field without nonlinear evolution is generated by taking the Gaussian random field and adding $\fnl=100$ following the definition of the primordial potential of the local type, Eqn~\ref{eq:local_potential}.
This field and the difference between two initial conditions are at the same level. The small difference in power might be due to the $\fnl=100$ effect on 2LPT, which we will explore in future studies.

\acknowledgments

We thank Anthony Challinor, Chihway Chang, Shi-Fan (Stephen) Chen, Neal Dalal, Simone Ferraro, Nick Gnedin, Sam Goldstein, Boryana Hadzhiyska, Colin Hill, Chris Hirata, Austin Joyce, Kazuya Koyama, Uroš Seljak, Román Scoccimarro, Ravi Sheth, Sergey Sibiryakov, Kendrick Smith, Jamie Sullivan, Frank van den Bosch, Eleonora Vanzan, Licia Verde, Matias Zaldarriaga for helpful discussions and comments. XC is supported by Future Investigators in
NASA Earth and Space Science and Technology (FINESST) grant
(award \#80NSSC21K2041). NP is supported in part by DOE DE-SC0017660 and NASA 80NSSC24M0021. DJE’s contributions were supported by U.S. Department of Energy grant DE-SC0007881, by the National Science Foundation under Cooperative Agreement PHY-2019786 (the NSF AI Institute for Artificial Intelligence and Fundamental Interactions, http://iaifi.org/), and as a Simons Foundation Investigator.

% This is the most common positions for acknowledgments. A macro is
% available to maintain the same layout and spelling of the heading.

% \paragraph{Note added.} This is also a good position for notes added
% after the paper has been written.

% The bibliography will probably be heavily edited during typesetting.
% We'll parse it and, using the arxiv number or the journal data, will
% query inspire, trying to verify the data (this will probalby spot
% eventual typos) and retrive the document DOI and eventual errata.
% We however suggest to always provide author, title and journal data:
% in short all the informations that clearly identify a document.

%\printbibliography
\bibliographystyle{JHEP}
\bibliography{ads-png-recon, png_phi2delta}

\end{document}